\documentclass[]{aa}
\usepackage[varg]{txfonts}
\usepackage{natbib,graphicx,subcaption,CJK}
\usepackage{hyperref}
\bibpunct{(}{)}{;}{a}{}{,}

\titlerunning{Concentrating small particles through the streaming instability}

\newcommand{\pc}{\textsc{Pencil Code}}
\newcommand{\phn}{\phantom{0}}

\makeatletter
\renewcommand*\aa@pageof{, page \thepage{} of \pageref*{LastPage}}
\makeatother

\begin{document}

\begin{CJK}{UTF8}{bkai}
\title{Concentrating small particles in protoplanetary disks\\
       through the streaming instability}
              
\author{C.-C.~Yang (楊朝欽) \and
        A.~Johansen \and
        D.~Carrera\thanks{Present address:
            Center for Exoplanets and Habitable Worlds,
            525~Davey Laboratory,
            The Pennsylvania State University,
            University Park, PA~16802, U.S.A.}}

\institute{Lund Observatory, Department of Astronomy and Theoretical Physics, Lund University,
           Box~43, SE-221\,00 Lund, Sweden\\
           \email{ccyang@astro.lu.se}}
\date{Received 21 November 2016 / Accepted 14 June 2017}

\abstract{
Laboratory experiments indicate that direct growth of silicate grains via mutual collisions can only produce particles up to roughly millimeters in size.
On the other hand, recent simulations of the streaming instability have shown that mm/cm-sized particles require an excessively high metallicity for dense filaments to emerge.
Using a numerical algorithm for stiff mutual drag force, we perform simulations of small particles with significantly higher resolutions and longer simulation times than in previous investigations.
We find that particles of dimensionless stopping time $\tau_\mathrm{s} = 10^{-2}$ and $10^{-3}$ -- representing mm- and cm-sized particles interior of the water ice line -- concentrate themselves via the streaming instability at a solid abundance of a few percent.
We thus revise a previously published critical solid abundance curve for the regime of $\tau_\mathrm{s} \ll 1$.
The solid density in the concentrated regions reaches values higher than the Roche density, indicating that direct collapse of particles down to mm sizes into planetesimals is possible.
Our results hence bridge the gap in particle size between direct dust growth limited by bouncing and the streaming instability.}


\maketitle
\section{Introduction}

In the core-accretion scenario of planet formation, planetary cores are assembled beginning with interstellar $\mu$m-sized dust grains \citep{vS69}.
This process of growing planetary cores covers more than 30 orders of magnitude in mass and more than 13 orders of magnitude in size, required to be completed within the 1--5 Myr lifetime of their natal protoplanetary disks \citep[e.g.,][]{HLL01,eM09}.
The course of planet formation is usually divided into different stages according to the size of the solids or planetary objects involved, and each stage has its own major difficulties \citep[e.g.,][]{PT06,BC14,TB14}.
One of these stages yet to be understood is the formation of kilometer-scale planetesimals \citep[we refer to e.g.,][and references therein]{JB14}.

A major obstacle to the formation of planetesimals is the ``radial-drift barrier.''
Solid particles marginally coupled to the gas via drag force drift radially inwards and are quickly removed from protoplanetary disks due to the gaseous head wind \citep{AHN76,sW77a}.
For example, the timescale for the radial drift of meter-sized boulders at $\sim$1 au of the minimum-mass solar nebula \citep[MMSN;][]{sW77b,cH81} is $\sim$100 yr, significantly shorter than the typical lifetime of protoplanetary disks.
In fact, solid particles of a wide range of sizes at various locations in disks suffer from radial drift \citep[e.g.,][]{BD07,aY10}.
One possibility to circumvent this barrier is to increase the collisional cross-section of the particles by collecting porous icy dust aggregates \citep{KT13}.
This process, however, may only operate outside the ice line and relies on the presence of large amounts of sub-micron monomers.
Therefore, some mechanism(s) to efficiently concentrate solid materials into high density so that direct gravitational collapse can occur appears to be required to form planetesimals.

Two types of mechanism that concentrate solid materials exist; one passive, one active.
The former includes long-lived local pressure maxima \citep{fW72,JYK09,BS14} or vortices \citep{BS95,BC99,KH14,wL14}, in which solids passively follow the underlying flow of gas by friction.
By contrast, the streaming instability discovered by \cite{YG05} is realized by the action-reaction pair of the drag force between solid particles and gas, with which the solids actively engage in the dynamics with the gas to spontaneously concentrate themselves into high densities \citep{JY07,BS10a,YJ14}.
As a result, not only does the dense filamentary structure of solids driven by the streaming instability have significantly reduced radial drift speeds and thus longer radial drift time-scales, but also planetesimals can form by direct gravitational collapse within these dense filaments \citep{JYL12,JM15,SA16,SA17,SYJ17}.

Nevertheless, it remains problematic for the streaming instability to drive planetesimal formation inside the ice line of protoplanetary disks \citep{DD14}.
Direct dust growth by coagulation of compact\footnote{The behavior of bouncing might only occur for dust aggregates of volume filling factor greater than $\sim$0.3 \citep{WT11} or $\sim$0.5 \citep{SK13}, as determined by numerical simulations.} silicate grains is limited to mm sizes, due to bouncing and fragmentation at collisions \citep{ZO10,BKE12}.
On the other hand, there exists a threshold in solid abundance only above which can the streaming instability drive solid concentration into high densities \citep{JYM09,BS10a}.
Using a suite of two-dimensional simulations, \cite{CJD15} found that this threshold abundance increases drastically with decreasing particle size.
For mm-sized particles inside the ice line of protoplanetary disks, they suggested that a solid-to-gas column density ratio of more than 10\% is required, an exceedingly high value in usual disk conditions.
Therefore, it seems that a significant gap between dust coagulation and the onset of planetesimal formation exists inside the ice line of young protoplanetary disks.

A few properties of the streaming instability are worth noticing, though, which may explain the exceedingly high critical solid abundance found for small particles.
First, even though the growth rate of the linear streaming instability for small particles is relatively small compared to that for large particles when the local solid-to-gas density ratio is low, it significantly overtakes that for large particles when the density ratio is more than the order of unity \citep{YJ07}.
Hence, the response of small particles to the streaming instability should be more prominent when the latter condition is reached.
However, the wavelengths of the fastest-growing modes are roughly proportional to the dimensionless stopping time $\tau_\mathrm{s} \equiv \Omega_\mathrm{K} t_\mathrm{s}$ and thus the size of the particles, where $\Omega_\mathrm{K}$ is the local Keplerian frequency and $t_\mathrm{s}$ is the stopping time characterizing the mutual drag force between the gas and the particles \citep{YJ07}.
These wavelengths are as small as $\sim$10$^{-4}H$ for particles with $\tau_\mathrm{s} \sim 10^{-3}$, where $H$ is the local scale height of the gas \citep{BS10b}.
Resolving these faster growing modes of the streaming instability remains challenging in current numerical simulations.
Although it remains unclear how these linear modes cooperate to determine the conditions for strong concentration of solid particles in the nonlinear saturation of the streaming instability \citep{YJ07,JBL11}, systematic resolution studies considering higher resolutions seem warranted.
Finally, in the nonlinear stage of the streaming instability, the timescales for sedimentation and radial drift of the particles should continue to be relevant for the system.
These timescales are inversely proportional to $\tau_\mathrm{s}$ when $\tau_\mathrm{s} \ll 1$, and hence a proportionately longer time may be required for small particles to concentrate themselves than for their large counterparts.

In spite of that, due to stringent time-step constraint of the stiff mutual drag force, it is numerically challenging to simulate the particle-gas system in question with a higher resolution and longer simulation time than was achieved by \cite{CJD15}.
To make such simulations feasible, we have devised a new algorithm in \cite{YJ16} to relieve this time-step constraint, and have demonstrated that the algorithm manages small particles and/or strong local solid concentration with satisfactory numerical convergence and accuracy.
Employing this technique in this work, we revisit \cite{CJD15} for the case of dimensionless stopping time $\tau_\mathrm{s} \ll 1$ with significantly higher resolutions and longer simulation times.

We indeed find these small particles can spontaneously concentrate themselves into high density at a much reduced solid abundance: $\sim$1--2\% for particles of $\tau_\mathrm{s} = 10^{-2}$ while $\sim$3--4\% for particles of $\tau_\mathrm{s} = 10^{-3}$.
In the following section, we briefly describe our methodology.
We then present the evolution of the system, focusing on the concentration of the solid particles, in both two-dimensional (2D; Sect.~\ref{S:2d}) and three-dimensional (3D; Sect.~\ref{S:3d}) models.
The implications for the formation of planetesimals are discussed in Sect.~\ref{S:impl}, where we revise the critical solid abundance of \cite{CJD15} with the results of this work.
We conclude with a short summary in Sect.~\ref{S:summary}.

\section{Methodology} \label{S:method}

As in our previous publication (\cite{YJ14}), we use the local-shearing-box approximation \citep{GL65} to simulate a system of gas and solid particles.
The computational domain is a small rectangular box at an arbitrary distance from the central star.
Its origin rotates around the star at the local Keplerian angular speed $\Omega_\mathrm{K}$ and its three axes constantly align in the radial, azimuthal, and vertical directions, respectively.
We consider isothermal, non-magnetized gas in a regular Eulerian grid with numerous Lagrangian solid particles in it.
The gas interacts with each of the particles via their mutual drag force, which is characterized by the stopping time $t_\mathrm{s}$ \citep{fW72,sW77a} or its dimensionless counterpart $\tau_\mathrm{s} \equiv \Omega_\mathrm{K} t_\mathrm{s}$ \citep{YG05}.
We include the linearized vertical gravity due to the central star on the particles but ignore it for the gas, since the computational domain considered in this work is so small compared to the vertical scale height of the gas that there is no appreciable vertical density stratification in the gas over the domain.
For simplicity, we assume that all particles have the same stopping time (we refer to the discussion in Sect.~\ref{S:impl}).
We also ignore collisions between and self-gravity of the particles in order to isolate the effects driven by the streaming instability.
The standard sheared periodic boundary conditions are imposed \citep{BN95,HGB95}, and we assume the vertical dimension is also periodic.

To evolve this system of gas and particles, we use the \pc\footnote{The \pc\ is publicly available at \url{http://pencil-code.nordita.org/}}, a high-order finite-difference simulation code for astrophysical fluids and particles \citep{BD02}.
It employs sixth-order centered differences for all spatial derivatives and a third-order Runge-Kutta method to integrate the system of equations.
To maintain numerical stability, hyper-diffusion operators are needed in each dynamical field variable for the gas, and we fix the mesh Reynolds number for these operators so that noise close to the Nyquist frequency is properly damped while the power over large dynamical range is preserved \citep{YK12}.
Furthermore, we use a sixth-order B-spline interpolation to integrate the shear advection terms \citep[c.f.,][]{JYK09}.
As mentioned above, the solid particles are modeled as Lagrangian (super-)particles, which have individual positions and velocities that are integrated in unison with the Runge--Kutta steps.
Their interaction with the Eulerian gas is achieved by the standard particle-mesh method \citep[e.g.,][]{HE88}.
To obtain high spatial accuracy, we choose the Triangular-Shaped-Cloud scheme for the particle-mesh interpolation and assignment \citep{YJ07}.

Since we focus our attention in this work on small particles with potentially strong local solid concentrations, the mutual drag force between the gas and the particles is stiff.
In this regard, we employ the numerical algorithm developed in \cite{YJ16} to relieve the stringent time-step constraint due to this stiffness of the drag force.
This algorithm intricately decomposes the globally coupled system of equations and makes it possible to integrate the equations on a cell-by-cell basis.
It uses the analytical solutions in each cell to achieve numerical stability with an arbitrary time step.
The momentum feedback from the particles to the gas cells is also an essential ingredient to expedite numerical convergence.

\begin{table*}
\caption{Model specifications\label{T:spec}}
\centering
\begin{tabular}{c c c c c}
\hline\hline
Dimensionless          & Solid         & Computational & Resolutions & Simulation Time\\
Stopping Time $\tau_\mathrm{s}$ & Abundance $Z$ & Domain        & ($H^{-1}$)  & ($P$)\\
\hline
\multicolumn{5}{c}{2D (radial-vertical) Models}\\
\hline
$10^{-2}$ & 0.01 & $0.2H \times 0.2H$ & 640, 1280, 2560       & 4000\\
          &      & $0.4H \times 0.4H$ & 320, 640, 1280        & 4000\\
          & 0.02 & $0.2H \times 0.2H$ & 640, 1280, 2560, 5120 & 1000\\
          &      & $0.4H \times 0.4H$ & 320, 640, 1280, 2560  & 1000\\
          & 0.04 & $0.2H \times 0.2H$ & 640, 1280, 2560       & 1000\\
          &      & $0.4H \times 0.4H$ & 320, 640, 1280        & 1000\\
\hline
$10^{-3}$ & 0.02 & $0.2H \times 0.2H$ & 640, 1280, 2560       & 1000\\
          &      & $0.4H \times 0.4H$ & 320, 640, 1280        & 1000\\
          & 0.03 & $0.2H \times 0.2H$ & 640, 1280, 2560       & 5000\\
          &      & $0.4H \times 0.4H$ & 320, 640, 1280        & 5000\\
          & 0.04 & $0.2H \times 0.2H$ & 640, 1280, 2560       & 2500\\
          &      & $0.4H \times 0.4H$ & 320, 640, 1280        & 2500\\
\hline
\multicolumn{5}{c}{3D Models}\\
\hline
$10^{-2}$ & 0.02 & $0.2H \times 0.2H \times 0.2H$ & 160, 320, 640 & 1000\\
\hline
$10^{-3}$ & 0.04 & $0.2H \times 0.2H \times 0.2H$ & 160, 320, 640 & 1000\\
\hline
\end{tabular}
\end{table*}

For comparison purposes, we investigate both 2D (radial-vertical) and 3D models.
The computational domain has a size of either 0.2$H$ or 0.4$H$ in each direction, where $H$ is the vertical scale height of the gas.
We consider solid particles of $\tau_\mathrm{s} = 10^{-2}$ and $\tau_\mathrm{s} = 10^{-3}$, which are approximately cm and mm in size (when their porosity is low) at $\sim$2.5 au in the primordial MMSN, respectively \citep[e.g.,][]{JB14}.
At later stages of disk evolution, these stopping times represent particles approximately ten times smaller \citep{BJ15}.
An external radial pressure gradient to the gas is imposed so that the azimuthal velocity of the gas is reduced by $\Delta v \equiv \Pi c_\mathrm{s} = 0.05c_\mathrm{s}$, a typical value in the inner region of the MMSN, where $c_\mathrm{s}$ is the isothermal speed of sound \citep{BS10a,BJ15}.
This external radial pressure gradient remains constant throughout the entire duration of each simulation.

The initial conditions are set as follows.
Given that it experiences no vertical gravity, the gas has a uniform density of $\rho_0$.
On the other hand, we allocate, on average, one particle per gas cell but randomly position the particles, such that the particle density distribution is vertically Gaussian with a constant scale height of 0.02$H$.
The initial scale height of particles is arbitrarily chosen in order to reduce the computing time for the initial sedimentation process, especially of small particles.
All the (super-)particles have the same mass, which is determined by the solid-to-gas mass ratio $Z \equiv \Sigma_{\mathrm{p},0} / \Sigma_{\mathrm{g},0}$, where $\Sigma_{\mathrm{p},0}$ and $\Sigma_{\mathrm{g},0} = \sqrt{2\pi}\rho_0 H$ are the initial uniform column densities of the particles and the gas, respectively \citep[][]{YJ14}.
With the density fields of the gas and the particles fixed, we apply the Nakagawa--Sekiya--Hayashi (\citeyear{NSH86}) equilibrium solutions to the horizontal velocities of the gas and the particles.
Their initial vertical velocities are set zero.

The atomic nature of the particles with a finite, constant mass limits the regions these particles can probe.
The mass of each particle is given by
\begin{equation}
m_{\mathrm{p}}
= \frac{\Sigma_{\mathrm{p},0}\Delta x \Delta y}{\bar{n}_\mathrm{p} N_z}
= \frac{\sqrt{2\pi}Z\rho_0 H\Delta x \Delta y}{\bar{n}_\mathrm{p} N_z},
\end{equation}
where $\Delta x$ and $\Delta y$ are the sizes of each cell in $x$- and $y$-directions, respectively, $\bar{n}_\mathrm{p}$ is the average number of particles per cell, and $N_z$ is the number of cells in $z$-direction.
This implies that the mass density each particle contributes to a cell is on the order of
\begin{equation}
\rho_{\mathrm{p},1}
= \frac{m_{\mathrm{p}}}{\Delta x \Delta y\Delta z}
= \frac{\sqrt{2\pi}Z\rho_0 H}{\bar{n}_\mathrm{p} L_z},
\end{equation}
where $\Delta z$ and $L_z$ are the sizes of each cell and the computational domain in $z$-direction, respectively.
Since each cell is sampled discretely by the particles, $\rho_{\mathrm{p},1}$ is also the minimum density of particles a simulation model can represent.
For $L_z = 0.2H$ and $Z = 0.02$, $\rho_{\mathrm{p},1} \simeq 0.25\rho_0 / \bar{n}_\mathrm{p}$.
At equilibrium state, the distribution of particles resembles a Gaussian function, and the maximum altitude a simulation model can reach is approximately
\begin{equation}
z_\mathrm{max}
= H_\mathrm{p}\sqrt{2\ln\left(\frac{Z\rho_0 H}{\rho_{\mathrm{p},1}H_\mathrm{p}}\right)}
= H_\mathrm{p}\sqrt{2\ln\left(\frac{\bar{n}_\mathrm{p} L_z}{\sqrt{2\pi}H_\mathrm{p}}\right)}  ,
\end{equation}
where $H_\mathrm{p}$ is the scale height of the particles.
For $\bar{n}_\mathrm{p} = 1$, $L_z = 0.2H$, and $H_\mathrm{p} = 0.015H$, $z_\mathrm{max} \simeq 1.8H_\mathrm{p}$, at which the density of particles is $\sim$19\% of that in the mid-plane.

Even though the layer of particles our simulation models can sample is restricted to $|z| \lesssim z_\mathrm{max}$, the dynamics of the particle-gas system near the mid-plane remains representative.
The particles in the high-altitude regions ($|z| \gtrsim z_\mathrm{max}$) only constitutes $\sim$7\% of the total mass of the particle layer in the above estimate.
The scale height of the particles is hence well approximated by the standard deviation of the vertical particle position $z_\mathrm{p}$:
\begin{equation}
H_\mathrm{p} \simeq \sqrt{\overline{z_\mathrm{p}^2} - \overline{z_\mathrm{p}}^2},
\end{equation}
which we use to identify $H_\mathrm{p}$ throughout this work.
We also note that near the mid-plane, the number of particles per cell is on the order of
\begin{equation}
n_\mathrm{p}(z = 0)
= \frac{Z\rho_0 H}{\rho_{\mathrm{p},1}H_\mathrm{p}}
= \frac{\bar{n}_\mathrm{p} L_z}{\sqrt{2\pi}H_\mathrm{p}}.
\end{equation}
For $\bar{n}_\mathrm{p} = 1$, $L_z = 0.2H$, and $H_\mathrm{p} = 0.015H$, $n_\mathrm{p}(0) \simeq 5$.
In Appendix~\ref{S:parres}, we vary the number of particles in our setup for comparison purposes.

\begin{figure*}
\centering
\includegraphics[width=17cm]{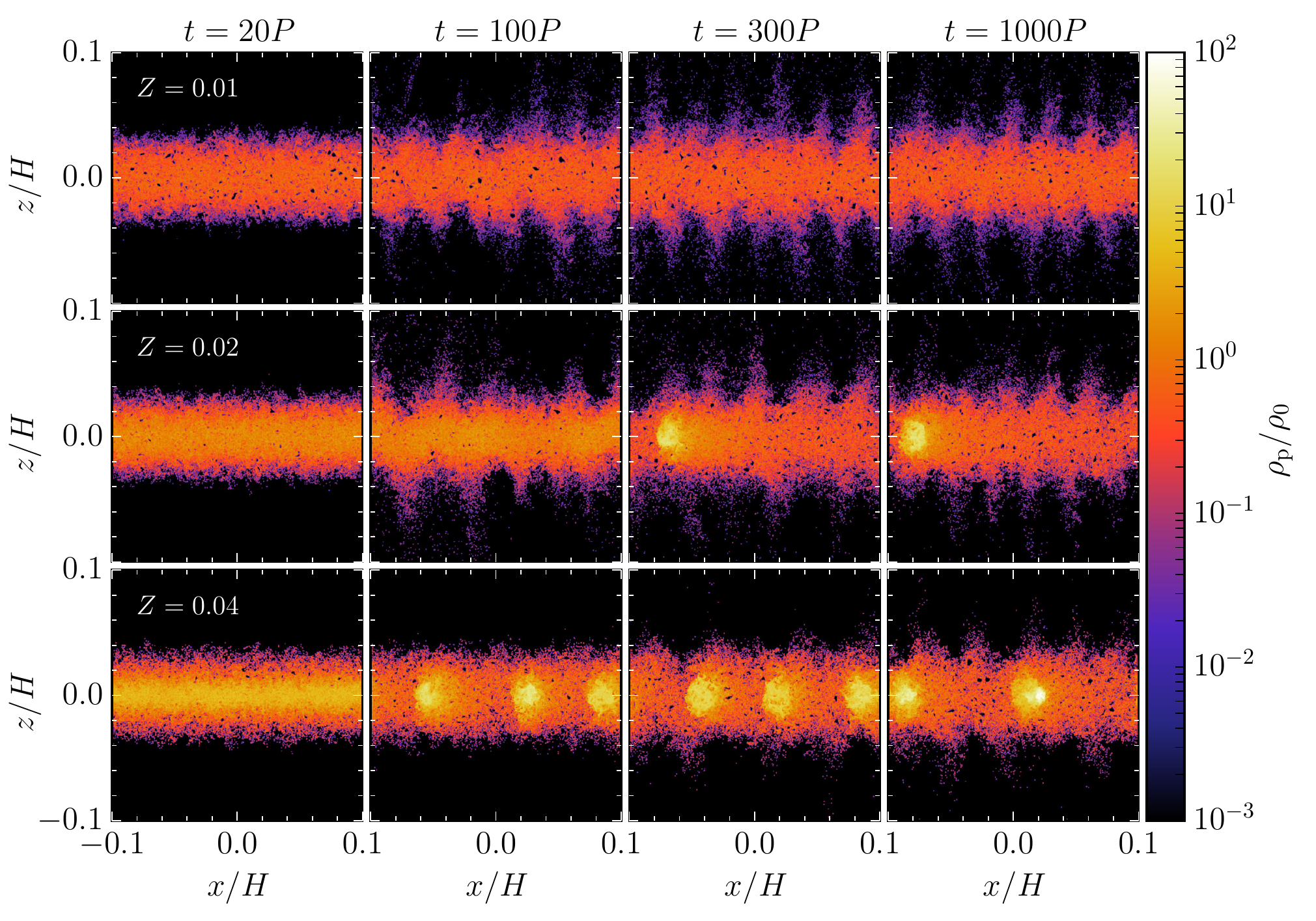}
\caption{Evolution of the particle density distribution for 2D models with particles of dimensionless stopping time $\tau_\mathrm{s} = 10^{-2}$.
The top, middle, and bottom panels show the models with an increasing solid abundance of $Z = 0.01$, 0.02, and 0.04, respectively, and the time $t$ in terms of the local orbital period $P$ increases from left to right.
The particle density $\rho_\mathrm{p}$ is measured with respect to the initial gas density in the mid-plane $\rho_0$, and the radial and vertical positions are expressed in terms of the vertical scale height of the gas $H$.
The models have a computational domain of 0.2$H$ $\times$ 0.2$H$ and a resolution of 2560$H^{-1}$.\label{F:evolt2}}
\end{figure*}

All the models investigated here are listed in Table~\ref{T:spec}.
For each set of model parameters, we evolve the system for at least 1000$P$, where $P = 2\pi / \Omega_\mathrm{K}$ is the local orbital period.
This is a significantly long simulation time compared to previous works in the literature, which is necessary to capture the timescale for the streaming instability to operate on and concentrate sedimented small particles, as discussed in detail in the following sections.

\section{Two-dimensional models} \label{S:2d}

In this section, we discuss the evolution of the density distribution of the particle layer found in our 2D models.
The system is axisymmetric in this case, with only radial and vertical variations.
We divide the discussion of the models by the dimensionless stopping time $\tau_\mathrm{s}$ of the particles.

\subsection{Particles of \texorpdfstring{$\tau_\mathrm{s} = 10^{-2}$}{taus=1E-2}} \label{SS:2dt2}

Figure~\ref{F:evolt2} shows the evolution of the particle layer from three of our 2D models with particles of $\tau_\mathrm{s} = 10^{-2}$.
They have the same computational domain of $0.2H \times 0.2H$ and resolution of 2560$H^{-1}$, but differ in their solid abundance $Z$: $Z = 0.01$, 0.02, and 0.04.

\begin{figure*}
\begin{center}
\includegraphics[width=17cm]{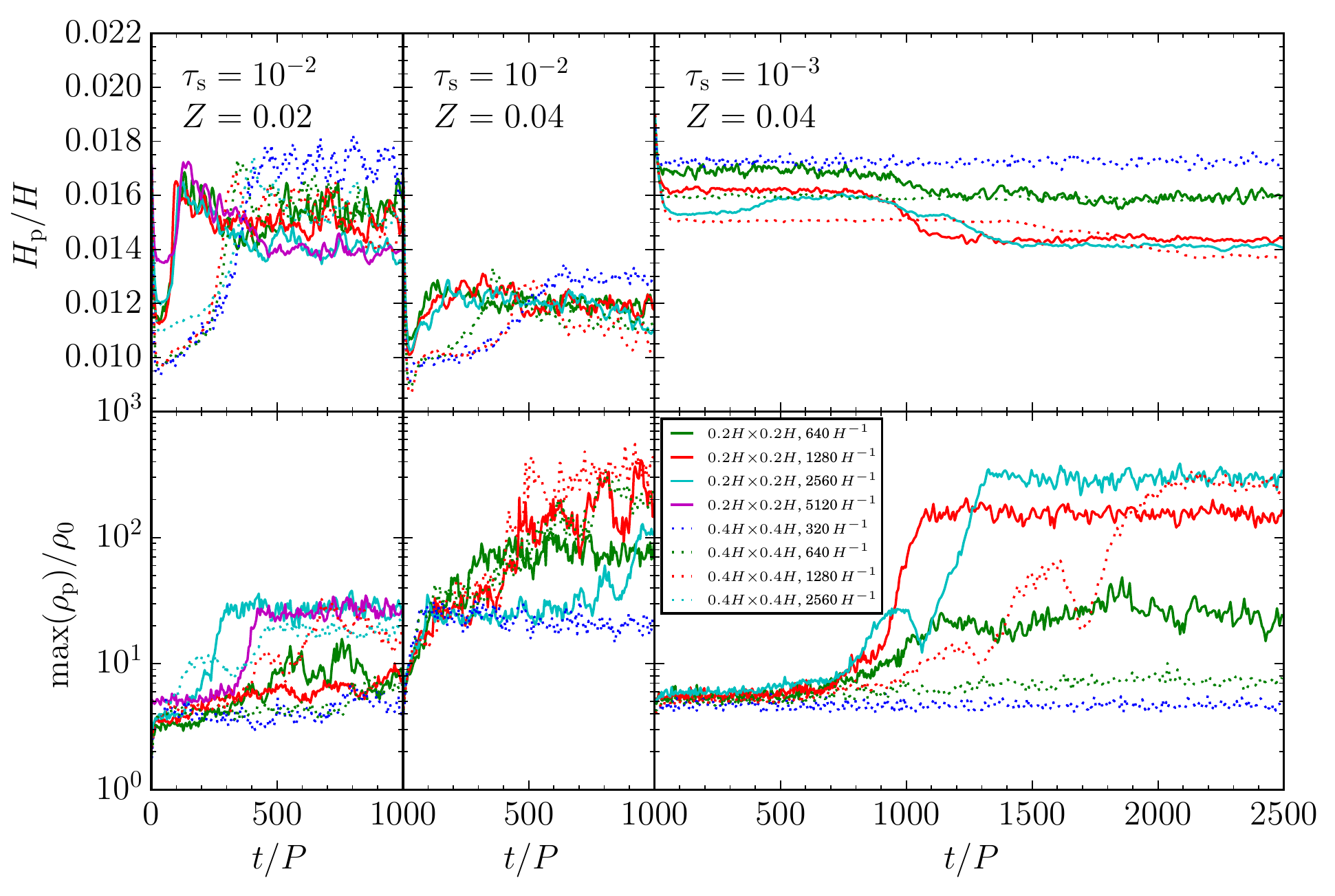}
\caption{Scale height of the particle layer $H_\mathrm{p}$ (top panels) and maximum of the local particle density $\rho_\mathrm{p}$ (bottom panels) as a function of time for all our 2D models that trigger strong concentration of solids at high resolutions.
Each column represents one set of dimensionless stopping time $\tau_\mathrm{s}$ and solid abundance $Z$.
The scale height $H_\mathrm{p}$, particle density $\rho_\mathrm{p}$, and time $t$ are normalized by the vertical scale height of the gas $H$, the initial mid-plane gas density $\rho_0$, and the local orbital period $P$, respectively.
The solid and dotted lines are for a computational domain of 0.2$H$ $\times$ 0.2$H$ and 0.4$H$ $\times$ 0.4$H$, respectively, and the resolutions are differentiated by different colors.
\label{F:ts2d}}
\end{center}
\end{figure*}

We first compare the cases of $Z = 0.01$ and $Z = 0.02$.
For $t \lesssim 100P$, both models follow similar evolution.
The particles continue their sedimentation toward the mid-plane during the first $\sim$20$P$, from the initial scale height of 0.02$H$ down to $\sim$0.013$H$ and $\sim$0.012$H$ for $Z = 0.01$ and 0.02, respectively (Fig.~\ref{F:ts2d}).
We note that the ($e$-folding) timescale for sedimentation is approximately $(2\pi\tau_\mathrm{s})^{-1}P$, which is $\sim$16$P$ for particles of $\tau_\mathrm{s} = 10^{-2}$.
In the meantime, random voids of various sizes driven by the streaming instability are developed  and they move around throughout the particle layer, which is the same phenomenon found in the unstratified simulations of saturated streaming turbulence of tightly-coupled particles \citep{JY07,YJ16}.
After $t \sim 20P$, the particles do not sediment anymore due to their random motion.
Furthermore, vertical oscillation of the particle layer starts to develop and the layer becomes increasingly more corrugated (Fig.~\ref{F:evolt2}).
In this manner, a small but appreciable fraction of the particles can be slung away from the mid-plane and move close to the vertical boundary.
As a result, the scale height of the particle layer gradually increases with time and reaches a level of $\sim$0.020$H$ and $\sim$0.016$H$ for $Z = 0.01$ and 0.02, respectively, when $t \sim 100P$ (Fig.~\ref{F:ts2d}).
A similar corrugation mode was also reported by \cite{LB15}, who suggested that it is driven by large-scale toroidal vortices.
However, we have not found evidence for such large-scale vortices in the particle layer in our models, due to complicated small-scale particle flow in the streaming turbulence.
Moreover, it remains unclear how boundary conditions affect this structure, which needs to be further investigated (R.~Li et al.\ 2016, in preparation; see also the discussion on the effect of other aspects of the model below).

\begin{figure*}
\centering
\begin{subfigure}{5.6cm}
   \centering
   \includegraphics[width=\textwidth]{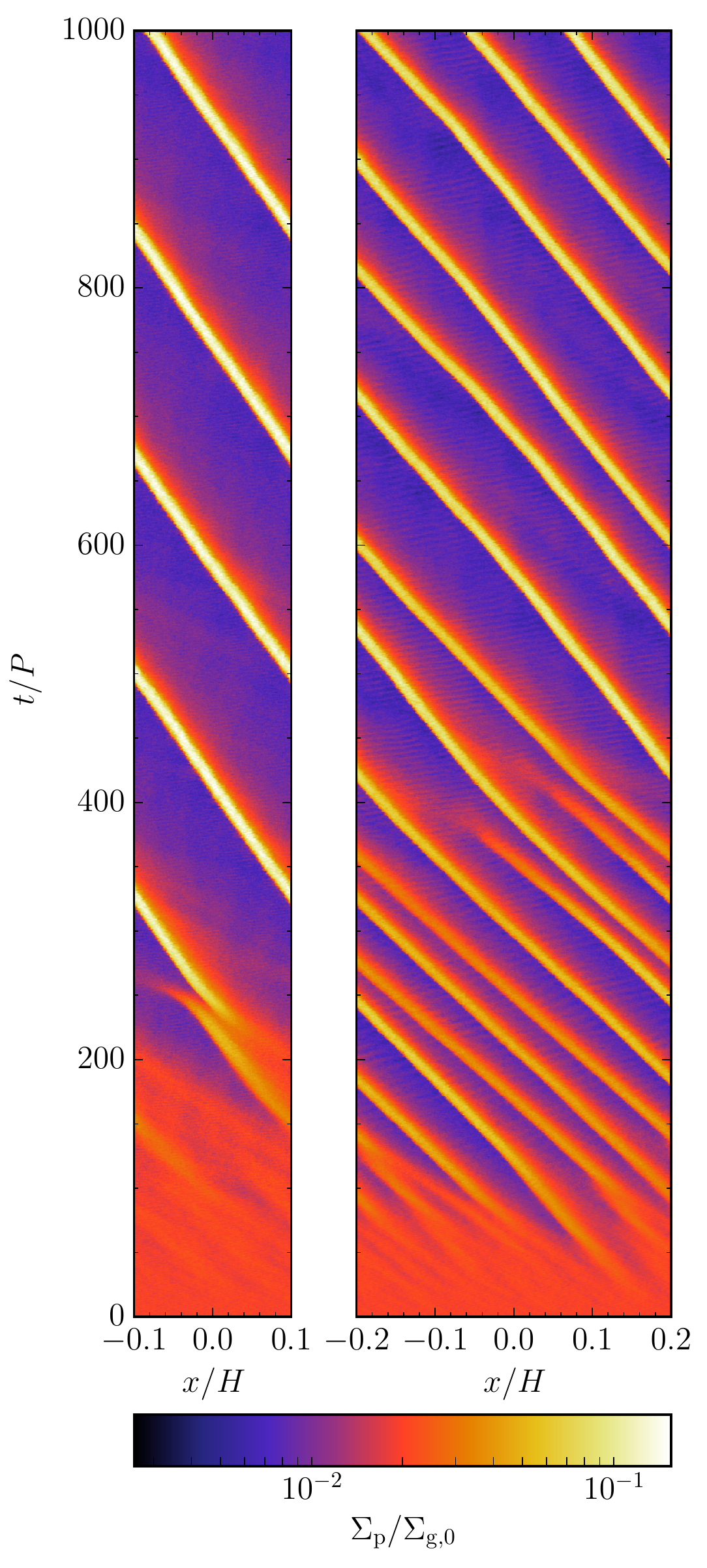}
   \caption{$\tau_\mathrm{s} = 10^{-2}, Z = 0.02, 2560H^{-1}$}
   \label{F:sigpt2}
\end{subfigure}
~
\begin{subfigure}{5.6cm}
   \centering
   \includegraphics[width=\textwidth]{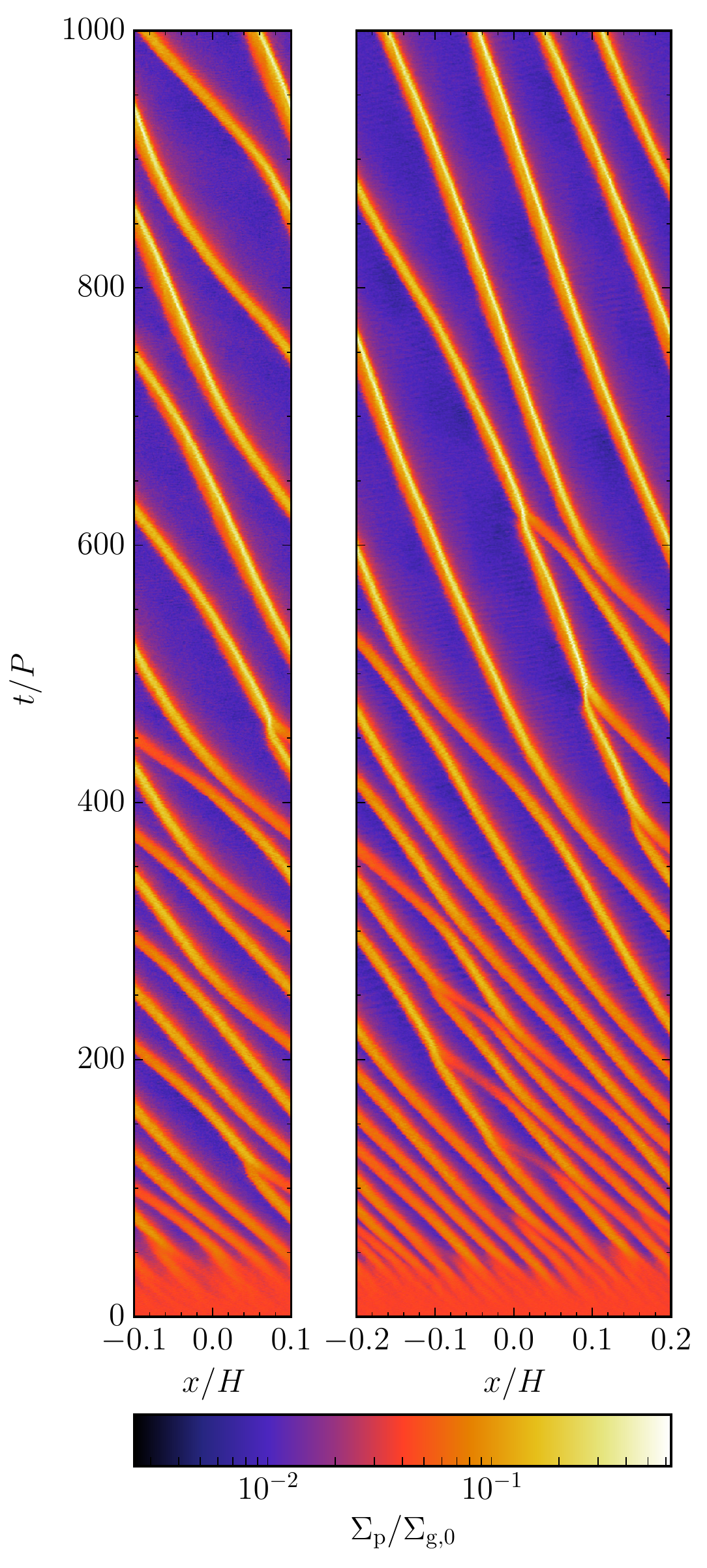}
   \caption{$\tau_\mathrm{s} = 10^{-2}, Z = 0.04, 1280H^{-1}$}
   \label{F:sigpt2z4}
\end{subfigure}
~
\begin{subfigure}{5.6cm}
   \centering
   \includegraphics[width=\textwidth]{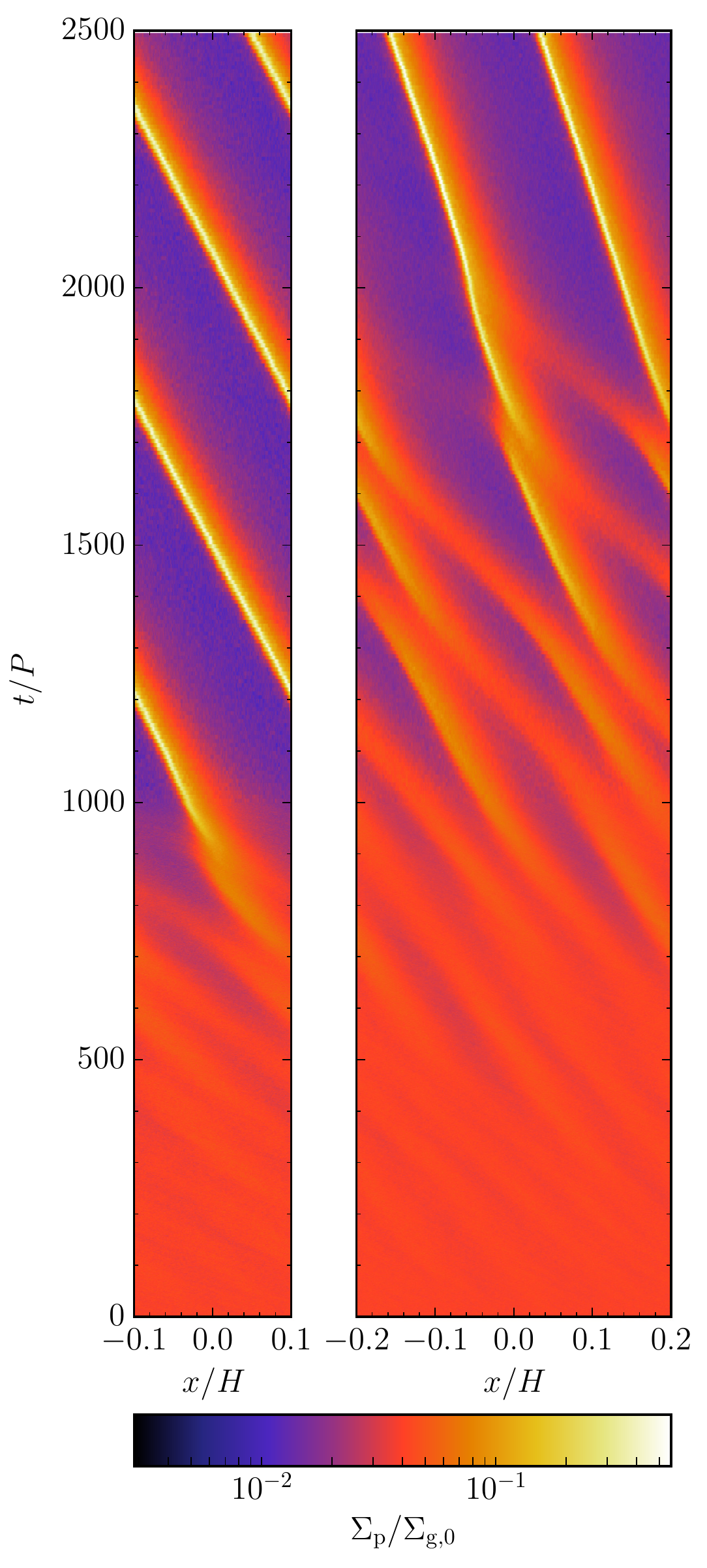}
   \caption{$\tau_\mathrm{s} = 10^{-3}, Z = 0.04, 1280H^{-1}$}
   \label{F:sigpt3}
\end{subfigure}
\caption{Evolution of the radial concentration of the particle layer for our 2D models that trigger strong concentration of solids.
Three cases of different dimensionless stopping time $\tau_\mathrm{s}$, solid abundance $Z$, and model resolution are presented.
The colors show the column density of the particles $\Sigma_\mathrm{p}$ as a function of radial position $x$ and time $t$, where $\Sigma_\mathrm{p}$, $x$, and $t$ are normalized by the initial gas column density $\Sigma_{\mathrm{g},0}$, the vertical scale height of the gas $H$, and the local orbital period $P$, respectively.
In each case, the left and right panels are from models with a computational domain of 0.2$H$ $\times$ 0.2$H$ and 0.4$H$ $\times$ 0.4$H$, respectively.
The timespan for case (c) is 2.5 times that for cases (a) and (b).\label{F:sigp2d}}
\end{figure*}

For $t > 100P$, the model with solid abundance $Z = 0.01$ maintains the same state up to $t = 4000P$, without any sign of further concentration of solid particles.
(We only show the state up to $t = 1000P$ in Fig.~\ref{F:evolt2}.)
By drastic contrast, however, the model with $Z = 0.02$ shows appreciable radial concentration of solids in the mid-plane over time, as shown in Fig.~\ref{F:evolt2}, the left panels of Fig.~\ref{F:ts2d}, and Fig.~\ref{F:sigpt2}.
A first dominant concentration begins at $t \sim 100P$ and reaches its peak at $t \sim 230P$.
A second even stronger concentration closely upstream appears soon afterward while the first is dispersed.
This second concentration reaches its saturation state at $t \sim 400P$ and then maintains its dominance and equilibrium up to the end of the simulation ($t = 1000P$).
At this state, the solid concentration attains a peak local density of approximately 30  times the background gas density (Fig.~\ref{F:ts2d}).

The left panels of Fig.~\ref{F:ts2d} show the evolution of the scale height and maximum local density of the particle layer for all of our 2D models with dimensionless stopping time $\tau_\mathrm{s} = 10^{-2}$ and solid abundance $Z = 0.02$.
For the first $\sim$50$P$, during which the streaming turbulence is established, it is seen that the higher the resolution, the earlier the establishment and the larger the local maximum density of solids result.
This is consistent with the fact that more and more faster growing modes of the linear streaming instability as well as smaller-scale structures in the particle layer are resolved \citep{YG05,YJ07}.
We note that even at our highest resolution, 5120$H^{-1}$, the fastest growing mode of the linear streaming instability remains well under-resolved, the wavelength of which is on the order of $10^{-4}H$ \citep[c.f.,][]{BS10b,YJ16}.
It is also seen that the higher the resolution, the higher the scale height of the particle layer in this initial stage.
On the other hand, the timescale to establish the corrugation of the particle layer does not seem to depend on resolution, but sensitively on computational domain.
As shown in Fig.~\ref{F:ts2d}, these timescales for our $0.2H\times0.2H$ and $0.4H\times0.4H$ models are $\sim$100$P$ and $\sim$400$P$, respectively.
Nevertheless, the degree of the vertical excitation of the particle layer is rather similar between the two computational domains and between different resolutions.

\begin{table*}
\caption{Saturation state of the 2D models\label{T:sat2d}}
\centering
\begin{tabular}{c c c c c c c c}
\hline\hline
Dimensions & Resolution & $t_\mathrm{sat}$ & $N_\mathrm{f}$ & $\langle H_\mathrm{p}\rangle$
    & $\langle\max(\rho_\mathrm{p})\rangle$
    & $\langle\Delta v_{\mathrm{p},x}\rangle$
    & $\langle\Delta v_{\mathrm{p},z}\rangle$\\
           & ($H^{-1}$) & ($P$)            &                & ($H$)
    & ($\rho_0$)
    & ($c_\mathrm{s}$)
    & ($c_\mathrm{s}$)\\
(1)        & (2)        & (3)              & (4)            & (5)
    & (6)
    & (7)     & (8)\\
\hline
\multicolumn{8}{c}{$\tau_\mathrm{s} = 10^{-2}$, $Z = 0.02$}\\
\hline
0.2$H$ $\times$ 0.2$H$
& \phn640 &    550 &   1--2 & 0.0155(4) & \phn9(2) & 0.0028(1) & 0.0028(1)\\
&    1280 &    800 &   1--2 & 0.0150(4) & \phn7(1) & 0.0027(1) & 0.0026(1)\\
&    2560 &    400 &   1    & 0.0141(3) &    28(4) & 0.0032(1) & 0.0033(2)\\
&    5120 &    450 &   1    & 0.0140(2) &    26(3) & 0.0038(2) & 0.0040(2)\\
0.4$H$ $\times$ 0.4$H$
& \phn320 &    500 &   1--5 & 0.0172(5) & \phn5(1) & 0.0024(1) & 0.0027(2)\\
& \phn640 &    950 &   2--3 & 0.0156(2) & \phn7(1) & 0.0022(1) & 0.0022(1)\\
&    1280 &    800 &   3    & 0.0144(3) &    18(4) & 0.0023(1) & 0.0023(1)\\
&    2560 &    500 &   3    & 0.0153(4) &    19(2) & 0.0029(1) & 0.0030(1)\\
\hline
\multicolumn{8}{c}{$\tau_\mathrm{s} = 10^{-2}$, $Z = 0.04$}\\
\hline
0.2$H$ $\times$ 0.2$H$
& \phn640 &    350 &   2    & 0.0120(2) & \phn79(15)    & 0.0029(1) & 0.0027(1)\\
&    1280 &    500 &   2    & 0.0119(2) &    176(73)    & 0.0027(1) & 0.0027(1)\\
&    2560 &    950 &   2    & 0.0110(1) &    114(9)\phn & 0.0027(1) & 0.0027(1)\\
0.4$H$ $\times$ 0.4$H$
& \phn320 &    600 &   5--6 & 0.0130(2) & \phn21(3)\phn & 0.0026(1) & 0.0025(1)\\
& \phn640 &    750 &   4    & 0.0113(2) &    225(45)    & 0.0023(1) & 0.0021(1)\\
&    1280 &    650 &   4    & 0.0110(4) &    339(70)    & 0.0021(1) & 0.0020(1)\\
\hline
\multicolumn{8}{c}{$\tau_\mathrm{s} = 10^{-3}$, $Z = 0.04$}\\
\hline
0.2$H$ $\times$ 0.2$H$
& \phn640 &   1200 &   1    & 0.0159(2) & \phn24(6)\phn & 0.0034(1) & 0.0033(1)\\
&    1280 &   1200 &   1    & 0.0144(1) &    156(18)    & 0.0028(1) & 0.0027(1)\\
&    2560 &   1500 &   1    & 0.0141(1) &    297(35)    & 0.0023(0) & 0.0023(0)\\
0.4$H$ $\times$ 0.4$H$
& \phn320 & \ldots & \ldots & 0.0172(1) &    \ldots        & 0.0032(1) & 0.0029(1)\\
& \phn640 &   2000 &   3--4 & 0.0159(0) & \phn\phn7(1)\phn & 0.0026(1) & 0.0024(1)\\
&    1280 &   2200 &   2    & 0.0137(0) &       266(23)    & 0.0021(0) & 0.0020(0)\\
\hline
\end{tabular}
\tablefoot{Columns:
(1)~Dimensions of the computational domain in terms of the gas scale height $H$;
(2)~Resolution in number of cells per $H$;
(3)~Estimated time to reach saturation in terms of the local orbital period $P$;
(4)~Number of major axisymmetric solid filaments;
(5)~Time-averaged scale height of the particles in $H$;
(6)~Time-averaged maximum local particle density in terms of the initial mid-plane gas density $\rho_0$;
(7)~Time-averaged radial velocity dispersion in terms of the isothermal speed of sound $c_\mathrm{s}$;
(8)~Time-averaged vertical velocity dispersion in $c_\mathrm{s}$.
The time averages are taken from $t = t_\mathrm{sat}$ to the end of the simulation with their standard deviation shown in parentheses.}
\end{table*}

The corresponding time averages of the properties shown in Fig.~\ref{F:ts2d} at the saturation state are summarized in Table~\ref{T:sat2d}.
There exists some critical model resolution above which significantly stronger solid concentration occurs; this critical resolution is around 512 grid points per dimension ($\sim$2560$H^{-1}$ and $\sim$1280$H^{-1}$ for the $0.2H\times0.2H$ and $0.4H\times0.4H$ models, respectively).
Convergence in the maximum local solid density at the saturated state for each computational domain is seen by comparing the models with the highest two resolutions.
The axisymmetric filaments of solids in the low-resolution models are less efficient in collecting particles upstream and easier to become dispersed, and new filaments can sporadically form in between the existing ones, competing for solid materials.
On the other hand, those in the high-resolution models are well concentrated and separated, leaving little material in between for any more filaments to form, as exemplified by Fig.~\ref{F:sigpt2}.
This dichotomy may be related with how well the scale height of the particle layer is resolved, which determines how many unstable modes of the linear streaming instability can operate in the simulation models.

Also listed in Table~\ref{T:sat2d} are the estimated time to reach saturation and the final number of dominant axisymmetric solid filaments for each computational domain and resolution.
There exists no obvious dependence of the saturation time on either dimensions or resolution, indicating the nonlinear and stochastic nature of the system.
It depends on the contentious interaction between the first generation of filaments before a stable configuration is secured (Fig.~\ref{F:sigp2d}).
Nevertheless, we establish that the timescale for particles of $\tau_\mathrm{s} = 10^{-2}$ to spontaneously concentrate themselves is on the order of 400$P$--950$P$.
Moreover, one and three major solid filaments tend to form in our $0.2H\times0.2H$ and $0.4H\times0.4H$ models, respectively.
This is consistent with the approximate 3:2 ratio in the maximum local solid density at the saturation state and indicates that the mass budget in the solid filaments in a $0.2H\times0.2H$ model can be overestimated \citep{SYJ17}.
It has been shown that for $\tau_\mathrm{s} \simeq 0.3$ particles, a model with at least 0.4$H$ on each side is needed to generate consistent density distribution of solids in the saturated state of the streaming instability, where multiple solid filaments can form (\citealt{YJ14}; R.~Li et al.\ 2016, in preparation).
Therefore, our measurement for the $0.4H\times0.4H$ models is likely to be more accurate.

\begin{figure*}[t]
\centering
\includegraphics[width=17cm]{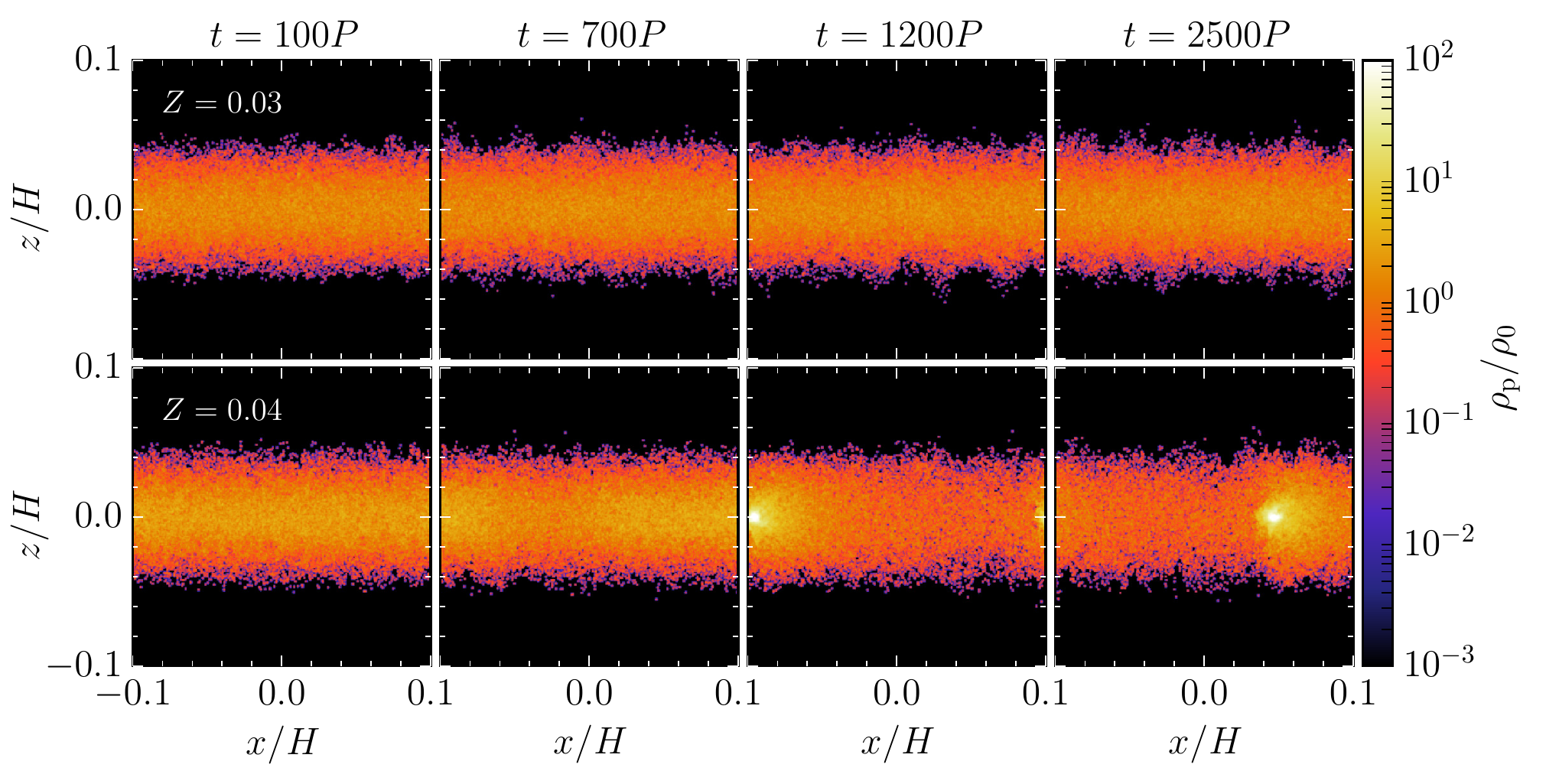}
\caption{Evolution of the particle density distribution for 2D models with particles of dimensionless stopping time $\tau_\mathrm{s} = 10^{-3}$.
The top and bottom panels show the models with a solid abundance of $Z = 0.03$ and $Z = 0.04$, respectively. Both of which have a computational domain of $0.2H \times 0.2H$ and a resolution of 1280\,$H^{-1}$.
The notations and the layout are otherwise the same as those in Fig.~\ref{F:evolt2}.\label{F:evolt3}}
\end{figure*}

Our models demonstrate that the critical solid abundance $Z_\mathrm{c}$ above which spontaneous strong concentration of particles of $\tau_\mathrm{s} = 10^{-2}$ is triggered lies in $0.01 < Z_\mathrm{c} < 0.02$; this is a somewhat smaller value than the range $0.02 < Z_\mathrm{c} < 0.03$ reported in \cite{CJD15}.
The reason for this discrepancy is that their models did not have enough resolution and simulation time.
\cite{CJD15} used a 128$\times$128 grid for a computational domain of $0.2H\times0.2H$.
With an initial solid abundance of $Z = 0.005$, they evolved the models for $\sim$50$P$ then artificially reduced the background gas density exponentially with time so that $Z = 0.08$ was attained at $t \sim 200P$, when the simulations ended.
However, as discussed above, we find that a grid of at least 512$\times$512 with a simulation time of at least 400$P$ is necessary to properly establish the saturation state of the solid concentration for particles of $\tau_\mathrm{s} = 10^{-2}$, a condition which was not satisfied by the models of \cite{CJD15}.
We show in the following subsection that this difference in measured values of critical solid abundance is significantly more drastic for particles of $\tau_\mathrm{s} = 10^{-3}$, which further indicates the importance of simulating small particles with high resolutions and long simulation times.

For comparison purposes, we also investigate the case of a solid abundance of $Z = 0.04$ for particles of $\tau_\mathrm{s} = 10^{-2}$.
The evolution of the particle scale height and the maximum local solid density for various resolutions and computational domains is shown in the middle panels of Fig.~\ref{F:ts2d}, and the properties at the saturation stage for these models are listed in Table~\ref{T:sat2d}.
Similar to the case of $Z = 0.02$, the streaming turbulence is established for the first $\sim$50$P$, and then the particle layer is slightly stirred up by the turbulence up to $t \sim 100P$--200$P$ for the $0.2H \times 0.2H$ models and $t \sim 400P$--600$P$ for the $0.4H \times 0.4H$ models.
However, the equilibrium scale height of the particle layer is appreciably less, at a level of $\sim$0.012$H$, and the corrugation mode is much less pronounced than those observed in the models for $Z = 0.02$, as shown in Fig.~\ref{F:evolt2}.
In general, higher solid abundance gives a more sedimented layer of particles.

On the other hand, conspicuous radial concentrations of solids occur as early as $t \sim 50P$, as shown in Fig.~\ref{F:evolt2}, the middle panels of Fig.~\ref{F:ts2d}, and Fig.~\ref{F:sigpt2z4}.
At this point, multiple concentrated filaments of solids with a density of a few tens of initial mid-plane gas density $\rho_0$ are already present, in contrast to the models for $Z = 0.02$.
In the following few hundreds of orbital periods, these filaments undergo a few merging events, forming denser filaments.
At the saturation stage when $t \gtrsim 500P$--900$P$, the $0.2H \times 0.2H$ models obtain a peak solid density of $\sim$100$\rho_0$--200$\rho_0$ with two dense filaments, while the $0.4H \times 0.4H$ models reach $\sim$200$\rho_0$--300$\rho_0$ with four dense filaments (Table~\ref{T:sat2d}).
Therefore, more solid filaments form with higher solid abundance $Z$, and the concentration of solids in these filaments appears to be a super linear function of $Z$.

\subsection{Particles of \texorpdfstring{$\tau_\mathrm{s} = 10^{-3}$}{taus=1E-3}} \label{SS:2dt3}

The evolution of the particle layer in our 2D models for particles of $\tau_\mathrm{s} = 10^{-3}$ is qualitatively similar to that for particles of $\tau_\mathrm{s} = 10^{-2}$ described in Sect.~\ref{SS:2dt2}, but there exist essential quantitative differences.
Figure~\ref{F:evolt3} shows such an evolution for two models with solid abundance $Z = 0.03$ and $Z= 0.04$, respectively, both of which have the same computational domain of $0.2H\times0.2H$ and resolution of 1280$H^{-1}$.
Similar to particles of $\tau_\mathrm{s} = 10^{-2}$, particles of $\tau_\mathrm{s} = 10^{-3}$ initially continue their sedimentation process until the streaming turbulence is well saturated and supporting the particle layer.
Given that the sedimentation timescale for particles of $\tau_\mathrm{s} = 10^{-3}$ is $\sim$160$P$, being one order of magnitude longer than that for particles of $\tau_\mathrm{s} = 10^{-2}$, the balance between sedimentation and turbulent excitation is not established until $t \sim 100P$--$200P$, as shown in the right panels of Fig.~\ref{F:ts2d}.
At this point, the scale height of the particle layer is $\sim$0.016$H$, somewhat higher than that for particles of $\tau_\mathrm{s} = 10^{-2}$.
We note that the largest voids of particles in the streaming turbulence are significantly smaller than those found in the models of particles of $\tau_\mathrm{s} = 10^{-2}$, which is consistent with the fact that the critical wavelength of the linear streaming instability decreases with decreasing $\tau_\mathrm{s}$ \citep{YG05,YJ07}.
The development of the corrugation of the particle layer can also be seen afterward.
However, the magnitude of this effect is not as strong as that for particles of $\tau_\mathrm{s} = 10^{-2}$.
The scale height of the particles slowly increases with time and reaches its peak at $t \sim 450P$, but the resulting excitation of the particle layer is barely noticeable (Fig.~\ref{F:evolt3}).

We find that particles of $\tau_\mathrm{s} = 10^{-3}$ with a solid abundance of either $Z = 0.02$ or $Z = 0.03$ remain in this statistically equilibrium state without strong concentration up to $t = 1000P$ and $t = 5000P$, respectively, in any of the model specifications listed in Table~\ref{T:spec}.
On the other hand, particles of $\tau_\mathrm{s} = 10^{-3}$ with $Z = 0.04$ can strongly concentrate themselves in the mid-plane later on (Fig.~\ref{F:evolt3}).
As shown in Fig.~\ref{F:sigpt3}, one axisymmetric solid filament begins to develop at $t \sim 700P$ and increases its concentration over time.
The concentration reaches its peak at $t \sim 1200P$ with a solid density of $\sim$160$\rho_0$, where $\rho_0$ is the initial gas density in the mid-plane (Fig.~\ref{F:ts2d}).
The filament is saturated and maintains its equilibrium state up to the end of the simulation ($t = 2500P$).

The right panels of Fig.~\ref{F:ts2d} shows the evolution of the scale height and maximum local density of the particle layer for all of our 2D models with dimensionless stopping time $\tau_\mathrm{s} = 10^{-3}$ and solid abundance $Z = 0.04$.
By comparing with the other models in Fig.~\ref{F:ts2d}, it is evident that although particles of either size follow a similar pattern of evolution, all the timescales involved with particles of $\tau_\mathrm{s} = 10^{-3}$ are significantly longer than those with particles of $\tau_\mathrm{s} = 10^{-2}$.
As is discussed above, these include the timescales for the initial balance between sedimentation and streaming turbulence to establish, for the corrugation mode to develop, and for the radial concentration of the particles in the mid-plane to reach saturation, the latter requiring more than 1000$P$.
We note also that in contrast to particles of $\tau_\mathrm{s} = 10^{-2}$, the initial scale height of the particles of $\tau_\mathrm{s} = 10^{-3}$ slightly decreases with increasing resolution.

Table~\ref{T:sat2d} lists the time-averaged properties of the particle layer at the final saturation stage for all of our 2D models.
Similar to particles of $\tau_\mathrm{s} = 10^{-2}$, there exists a critical resolution above which strong concentration of particles of $\tau_\mathrm{s} = 10^{-3}$ occurs; this resolution is $\sim$1280$H^{-1}$ for both $0.2H\times0.2H$ and $0.4H\times0.4H$ models.
We observe that the timescale for particles of $\tau_\mathrm{s} = 10^{-3}$ to spontaneously concentrate themselves into dense axisymmetric filaments is on the order of 1000$P$--2000$P$.
One and two major solid filaments tend to form in our $0.2H\times0.2H$ and $0.4H\times0.4H$ models, respectively.
The maximum local solid density in these filaments can be as high as $\sim$300$\rho_0$.
This is one order of magnitude higher than what is achieved by particles of $\tau_\mathrm{s} = 10^{-2}$ with an abundance of $Z = 0.02$.
It becomes comparable when either type of particles has $Z = 0.04$, but the former have relatively stronger concentration due to the smaller number of filaments formed.
Finally, the scale height of the particles in these models is also noticeably reduced when strong concentration occurs, which is due to further sedimentation of particles in the dense filaments.

Our models indicate that the critical solid abundance $Z_\mathrm{c}$ needed to trigger strong concentration of particles of $\tau_\mathrm{s} = 10^{-3}$ lies in the range $0.03 < Z_\mathrm{c} < 0.04$.
By contrast, \cite{CJD15} did not find any sign of significant concentration for solid abundances up to $Z \sim 0.2$.
It should be clear by now that the reason for this discrepancy is due to the long timescale ($\gtrsim$1000$P$) and high model resolution ($\gtrsim$1280$H^{-1}$) required to realize the process.
The 2D models in \cite{CJD15} had a resolution of 640$H^{-1}$ and a simulation time of $\sim$200$P$.
As shown in Fig.~\ref{F:ts2d}, only an equilibrated streaming turbulence can be observed during this period.

\begin{figure*}
\centering
\includegraphics[width=17cm]{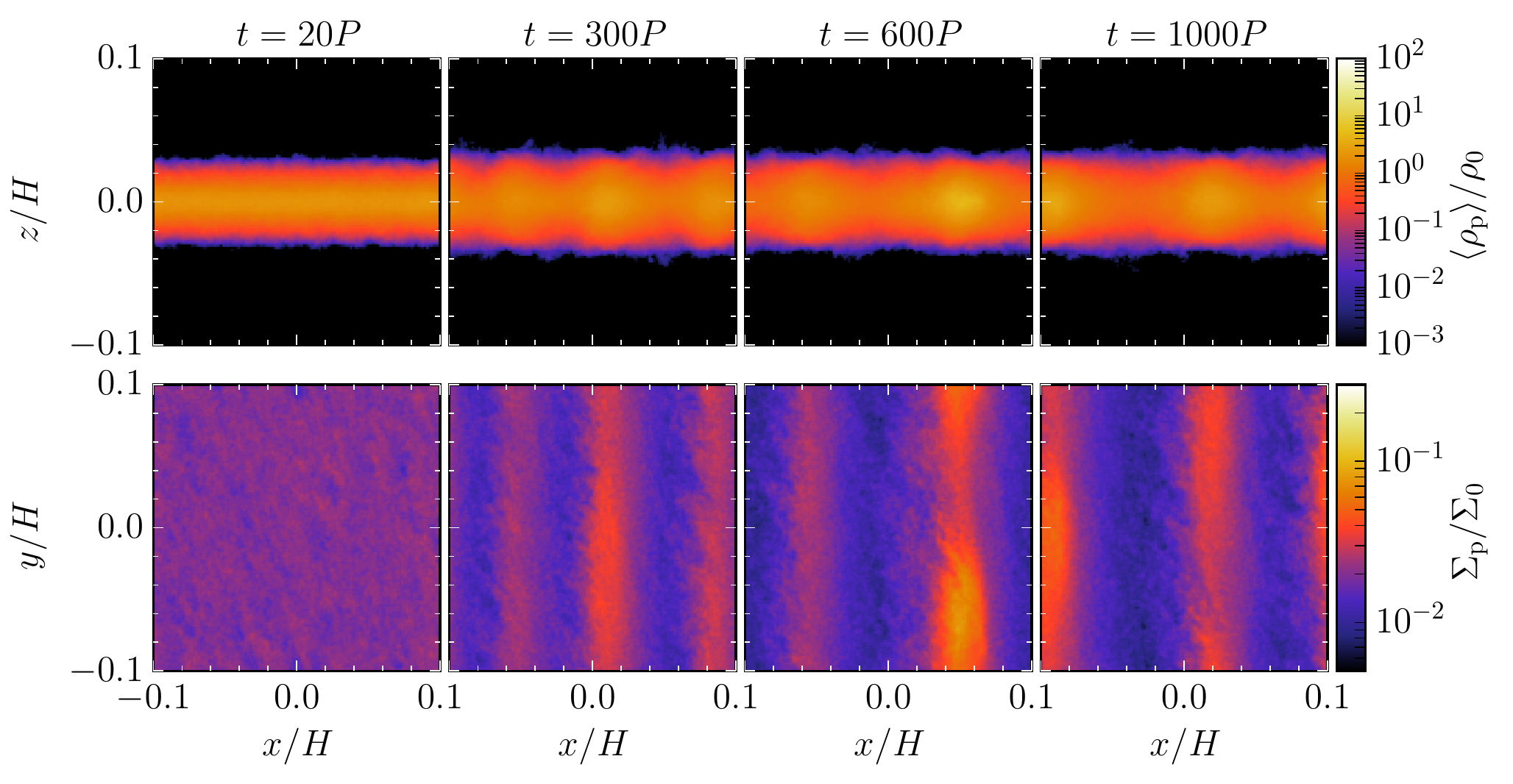}
\caption{Side and top views of the particle layer for a 3D model with particles of $\tau_\mathrm{s} = 10^{-2}$ and a solid abundance of $Z = 0.02$.
The time $t$ in terms of the local orbital period $P$ increases from left to right.
The top panels show the azimuthal average of the particle density $\rho_\mathrm{p}$ with respect to the initial gas density in the mid-plane $\rho_0$, while the bottom panels show the column density of solids $\Sigma_\mathrm{p}$ with respect to the initial column density of gas $\Sigma_0$.
The coordinates are normalized by the vertical scale height of the gas $H$.
The model has a computational domain of 0.2$H$ on each side and a resolution of 640$H^{-1}$.\label{F:evol3dt2}}
\end{figure*}

\begin{figure*}
\centering
\includegraphics[width=16.75cm]{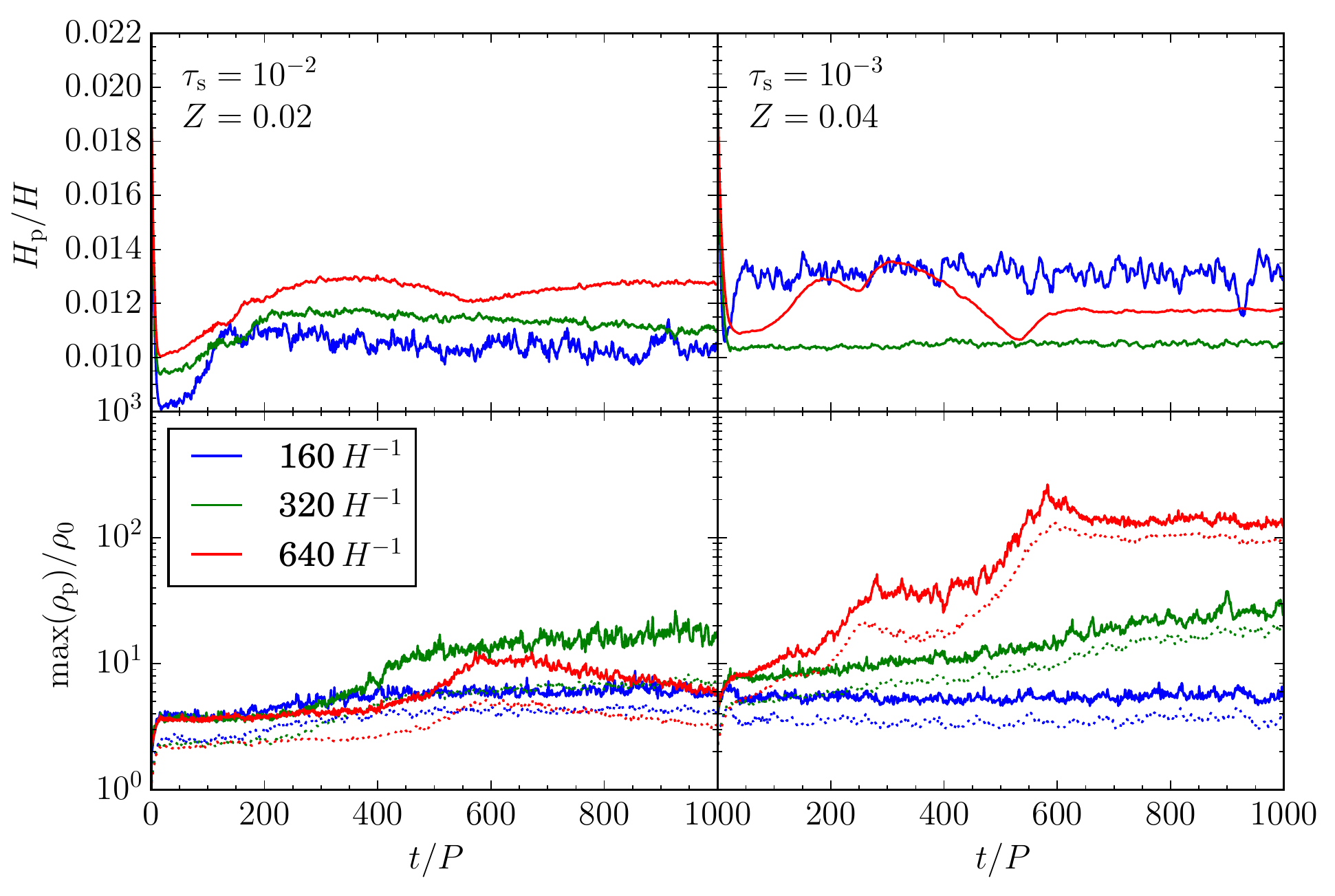}
\caption{Scale height of the particle layer $H_\mathrm{p}$ (top panels) and maximum of the local particle density $\rho_\mathrm{p}$ (bottom panels) as a function of time for all our 3D models.
Each column represents one set of dimensionless stopping time $\tau_\mathrm{s}$ and solid abundance $Z$.
All the models have a computational domain of 0.2$H$ on each side.
Different line colors represent different resolutions.
The dotted lines are obtained by first averaging over the azimuthal direction before taking the maximum of $\rho_\mathrm{p}$.
\label{F:ts3d}}
\end{figure*}

As a final remark, Table~\ref{T:sat2d} also lists the time-averaged radial and vertical velocity dispersions of the solid particles in the saturated stage of our 2D models with significant concentration.
They are in general about 0.2--0.3\% of the  speed of sound, irrespective of the stopping time and solid abundance.
However, we note that being an ensemble average of all the particles, the measured velocity dispersions do not reveal their spatial variation, and tend to be biased towards less dense regions.
Moreover, as presented in Appendix~\ref{S:rpg}, dynamically significant perturbation in the gas density may occur when the dimensionless stopping time $\tau_\mathrm{s} \ll 1$ and the local solid-to-gas density ratio $\epsilon \gg 1$.
Detailed analysis of the density and velocity structure in these models and the corresponding particle-gas dynamics will be the target of a future investigation.

\section{Three-dimensional models} \label{S:3d}

In this section, we use 3D models to study the effect of the azimuthal dimension on the spontaneous concentration of small solid particles.
We focus on a computational domain of 0.2 times gas scale height $H$ on each side and a solid abundance that can trigger significant solid concentration in the 2D models reported in Sect.~\ref{S:2d}.
We note that \cite{JY07} found that the morphological features of the streaming turbulence for tightly coupled particles in unstratified models are not axisymmetric, which can lead to stronger density fluctuations.
By comparing each 3D model with its 2D counterpart, we find how the azimuthal dimension can change the properties of a strictly axisymmetric distribution of sedimented solids.

\subsection{Particles of \texorpdfstring{$\tau_\mathrm{s} = 10^{-2}$}{taus=1E-2}} \label{SS:3dt2}

Figure~\ref{F:evol3dt2} shows the evolution of the particle layer for particles of dimensionless stopping time $\tau_\mathrm{s} = 10^{-2}$ and a solid abundance of $Z = 0.02$, at a resolution of 640$H^{-1}$.
Similar to its 2D counterpart, the particles obtain their balance between sedimentation and streaming turbulence within $\sim$20$P$, where $P$ is the local orbital period, as also shown in the left panels of Fig.~\ref{F:ts3d}.
The resulting scale height of particles is $\sim$0.010$H$, slightly lower than that in the 2D model.
From the top view of the particle disk, it can be seen that the streaming turbulence is indeed non-axisymmetric, with small-scale filamentary structures driven by background Keplerian shear.
This is consistent with what was found in unstratified models for tightly coupled particles \citep{JY07}.
From $t \sim 20P$ to $\sim$200$P$, the particles are gradually excited in the vertical direction, and the scale height slightly increases to the level of $\sim$0.012$H$.
In contrast to the corrugation mode found in 2D models, however, the excitation of the particles in the 3D models is not regular in the radial direction, and the level of the excitation is appreciably less.

\begin{figure}
\centering
\includegraphics[width=\columnwidth]{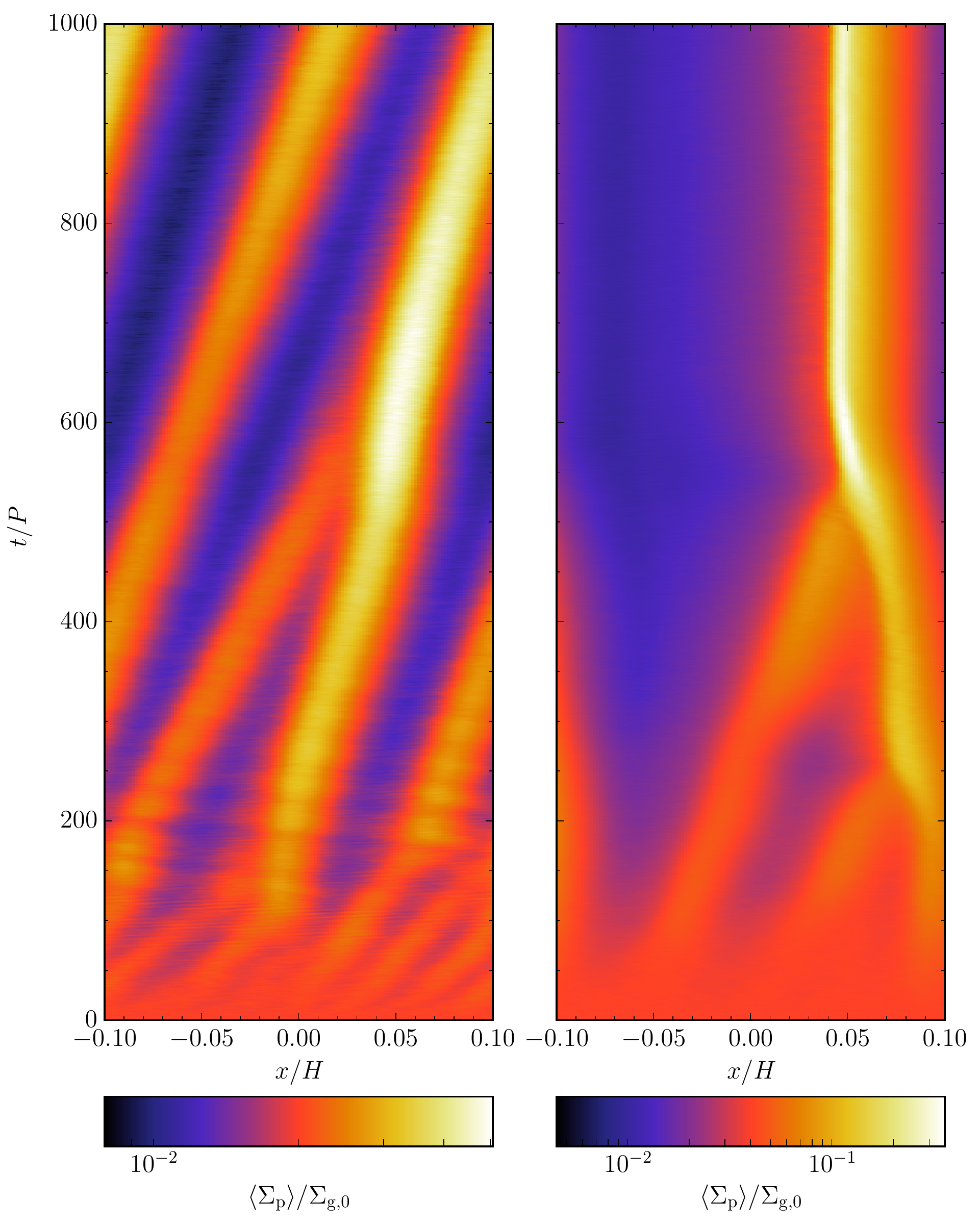}
\caption{Evolution of the radial concentration of the particle layer for two 3D models.
The colors show the azimuthal average of the column density of the particles $\Sigma_\mathrm{p}$ as a function of radial position $x$ and time $t$, where $\Sigma_\mathrm{p}$, $x$, and $t$ are normalized by the initial gas column density $\Sigma_{\mathrm{g},0}$, the vertical scale height of the gas $H$, and the local orbital period $P$, respectively.
The left panel has particles of dimensionless stopping time $\tau_\mathrm{s} = 10^{-2}$ and a solid abundance of $Z = 0.02$, while the right panel has $\tau_\mathrm{s} = 10^{-3}$ and $Z = 0.04$.
Both models have the same computational domain of 0.2$H$ on each side and resolution of 640$H^{-1}$.\label{F:sigp3d}}
\end{figure}

In the meantime, multiple radial concentrations of solids near the mid-plane begin to appear, as shown in Figs.~\ref{F:evol3dt2} and~\ref{F:sigp3d}.
As the solids continue to accumulate, the filamentary structures become more and more aligned in the azimuthal direction.
Interestingly, though, these filaments migrate radially outwards, contrary to what is seen in the 2D models.
Similar behavior was also seen in the 3D models conducted by \cite{CJD15}, who used an explicit method to integrate the mutual drag force.
We present a more detailed analysis of this behavior in Appendix~\ref{S:nonaxi}.
In spite of this, the maximum local density of solids in these filaments reaches approximately ten times the initial gas density in the mid-plane $\rho_0$ at $t \sim 600P$ (Fig.~\ref{F:ts3d}), a similar level and timescale observed in the 2D model of the same resolution and dimensions (Fig.~\ref{F:ts2d}).
At the end of the simulation when $t = 1000P$, the system has not yet obtained a statistically steady state.
We also note that the resolution we have achieved here remains lower than the critical resolution required to trigger significant concentration in the corresponding 2D models (Sect.~\ref{SS:2dt2}).

Also shown in the left panels of Fig.~\ref{F:ts3d} are the evolution of the scale height of the particle layer and the maximum local solid density for otherwise the same models but with two lower resolutions.
Similar to the trend found in the 2D models for particles of $\tau_\mathrm{s} = 10^{-2}$ (Sect.~\ref{SS:2dt2}), the equilibrium scale height increases slightly with increasing resolution at $t \sim 20P$, and all the models undergo moderate vertical excitation of particles during $t \sim 20P$ to $\sim$150$P$.
The two lower-resolution models, however, only drive one axisymmetric filamentary concentration of particles during $t \sim 200P$ to 1000$P$.
Because of this, the model with a resolution of 320$H^{-1}$ reaches a higher local solid density of $\sim$20$\rho_0$ in the filament than the model with a resolution of 640$H^{-1}$.
It remains to be investigated whether or not a convergent saturation state can be achieved with even higher-resolution models.

\subsection{Particles of \texorpdfstring{$\tau_\mathrm{s} = 10^{-3}$}{taus=1E-3}}

\begin{figure*}
\centering
\includegraphics[width=17cm]{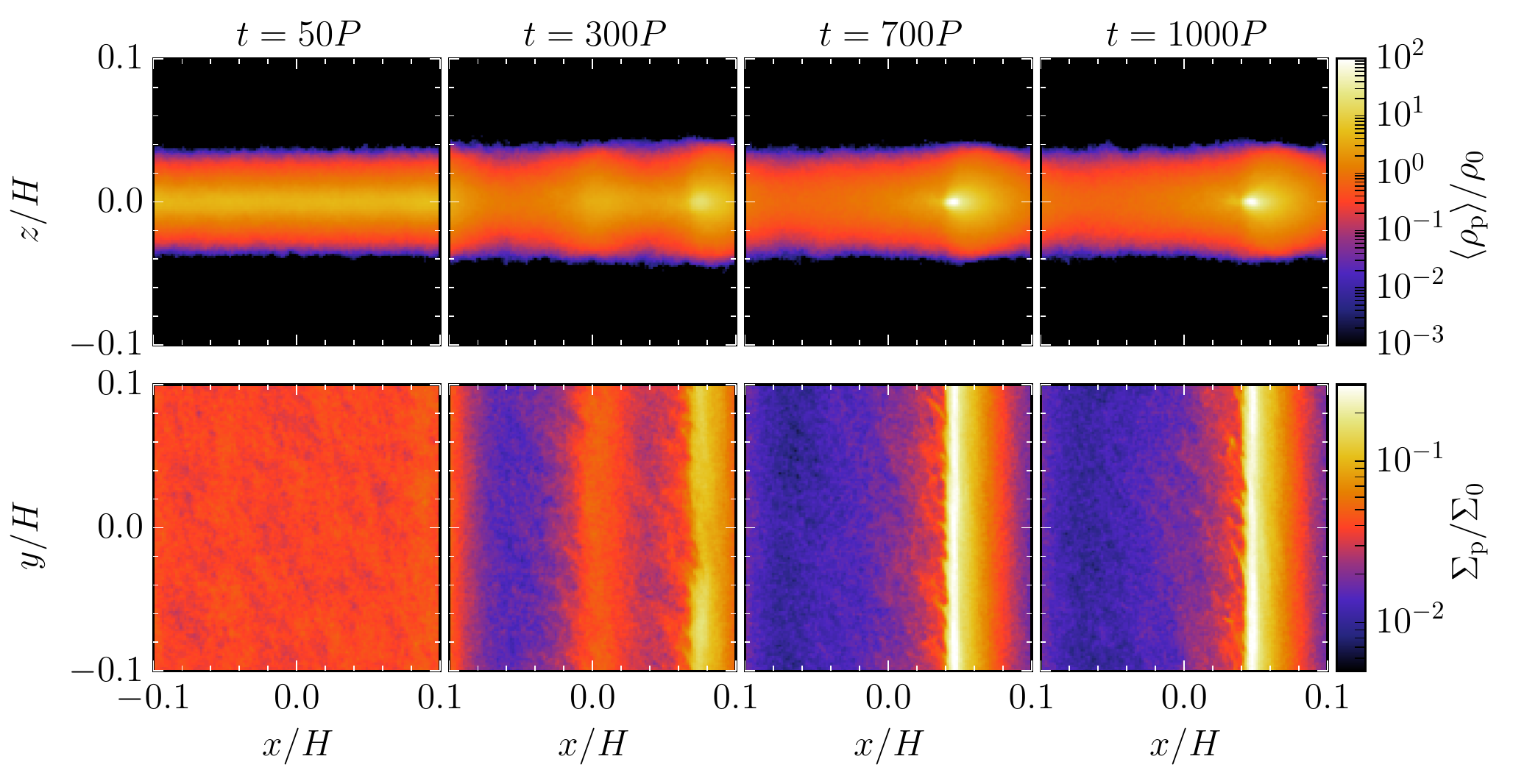}
\caption{Side and top views of the particle layer for a 3D model with particles of $\tau_\mathrm{s} = 10^{-3}$ and a solid abundance of $Z = 0.04$.
The notations, layout, and model specification are otherwise the same as those in Fig.~\ref{F:evol3dt2}.\label{F:evol3dt3}}
\end{figure*}

Figure~\ref{F:evol3dt3} shows the evolution of the particle layer for particles of dimensionless stopping time $\tau_\mathrm{s} = 10^{-3}$ and a solid abundance of $Z = 0.04$, at a resolution of 640$H^{-1}$.
Similar to its 2D counterpart at the same resolution, the balance between sedimentation and streaming turbulence is attained at $t \sim 50P$, as also shown in the right panels of Fig.~\ref{F:ts3d}.
The scale height of the particle layer at this point, though, is $\sim$0.011$H$, which is appreciably less than $\sim$0.017$H$ maintained by the 2D model (Fig.~\ref{F:ts2d}).
Nevertheless, the particles are gradually excited to a level of scale height $\sim$0.013$H$ over the period of $50P \lesssim t \lesssim 200P$.

As shown in the right panel of Fig.~\ref{F:sigp3d} as well as Fig.~\ref{F:evol3dt3}, three filaments of concentrated solids begin to develop at $t \sim 100P$.
Similar to the 3D model for particles of $\tau_\mathrm{s} = 10^{-2}$ at the same resolution of 640$H^{-1}$ (Sect.~\ref{SS:3dt2}), two of the three filaments migrate outwards.
They eventually merge with the third at $t \sim 250P$ and $t \sim 550P$, respectively.
As a result, the merged filament of solids becomes so dense that it virtually stops any further radial drift, and maintains its state up to the end of the simulation at $t = 1000P$.
The maximum local solid density reaches the order of $10^2\rho_0$ and the average scale height of the particles slightly decreases due to further sedimentation in the dense filament, as shown in the right panels of Fig.~\ref{F:ts3d}.
This concentration of solids is at the same level of what is obtained by the corresponding 2D models, albeit at a lower resolution than the critical one required by the latter (Sect.~\ref{SS:2dt3}).

The right panels of Fig.~\ref{F:ts3d} show the scale height of the particle layer and the maximum local solid density as a function of time for our 3D models with particles of $\tau_\mathrm{s} = 10^{-3}$ and $Z = 0.04$ at three different resolutions.
The model with the resolution of 160$H^{-1}$ does not show any sign of significant concentration of solids over the course of the simulation.
At the resolution of 320$H^{-1}$, accumulation of solids into one broad filament appears, and the density of the filament gradually and steadily increases with time, reaching $\sim$30$\rho_0$ at $t = 1000P$.
As discussed above, the model with the highest resolution of 640$H^{-1}$ shows strong concentration of solids and the system reaches its final saturated state at $t \sim 650P$, which is considerably less than the timescale of $\sim$1000$P$ required by its 2D counterpart (Fig.~\ref{F:ts2d} and Table~\ref{T:sat2d}).

\section{Implications for planetesimal formation} \label{S:impl}

The 2D and 3D models we present in Sects.~\ref{S:2d} and~\ref{S:3d} indicate that significant spontaneous concentration of solids in protoplanetary disks can occur at a considerably less solid abundance than reported in \cite{CJD15}, especially for particles as small as $\tau_\mathrm{s} = 10^{-3}$.
It is a matter of timescale and resolution for the process to operate.
For particles of $\tau_\mathrm{s} = 10^{-2}$ and those of $\tau_\mathrm{s} = 10^{-3}$, we identify that timescales of 400$P$--1000$P$ and 600$P$--2000$P$ are required, respectively.
We note that even though a considerably longer timescale is necessary for small particles to concentrate themselves, their radial drift timescale is also proportionately longer due to their small stopping times.
The radial drift timescale is given by \citep{AHN76}:
\begin{align}
t_\mathrm{drift}
&= \left(\frac{1 + \tau_\mathrm{s}^2}{4\pi\Pi\tau_\mathrm{s}}\right)
    \left(\frac{H}{R}\right)^{-1}P\\
&\simeq 3\times10^4 \left(\frac{\tau_\mathrm{s}}{10^{-3}}\right)^{-1}
                    \left(\frac{\Pi}{0.05}\right)^{-1}
                    \left(\frac{H / R}{0.05}\right)^{-1} P,
\end{align}
for $\tau_\mathrm{s} \ll 1$, where $\Pi \equiv \Delta v / c_\mathrm{s}$ and $\Delta v$ is the reduction in gas velocity due to radial pressure gradient.
The timescales for $\tau_\mathrm{s} = 10^{-2}$ and $\tau_\mathrm{s} = 10^{-3}$ are thus one order of magnitude longer than those required for driving radial concentration of solids by the streaming turbulence.
Therefore, these small particles do not suffer from the radial-drift barrier any worse than large particles.
Moreover, the radial-drift barrier is further alleviated once significant concentration of solids is established due to the back reaction of the drag force from the solids to the gas \citep{NSH86,JY07}.

\begin{figure}
\centering
\includegraphics[width=\columnwidth]{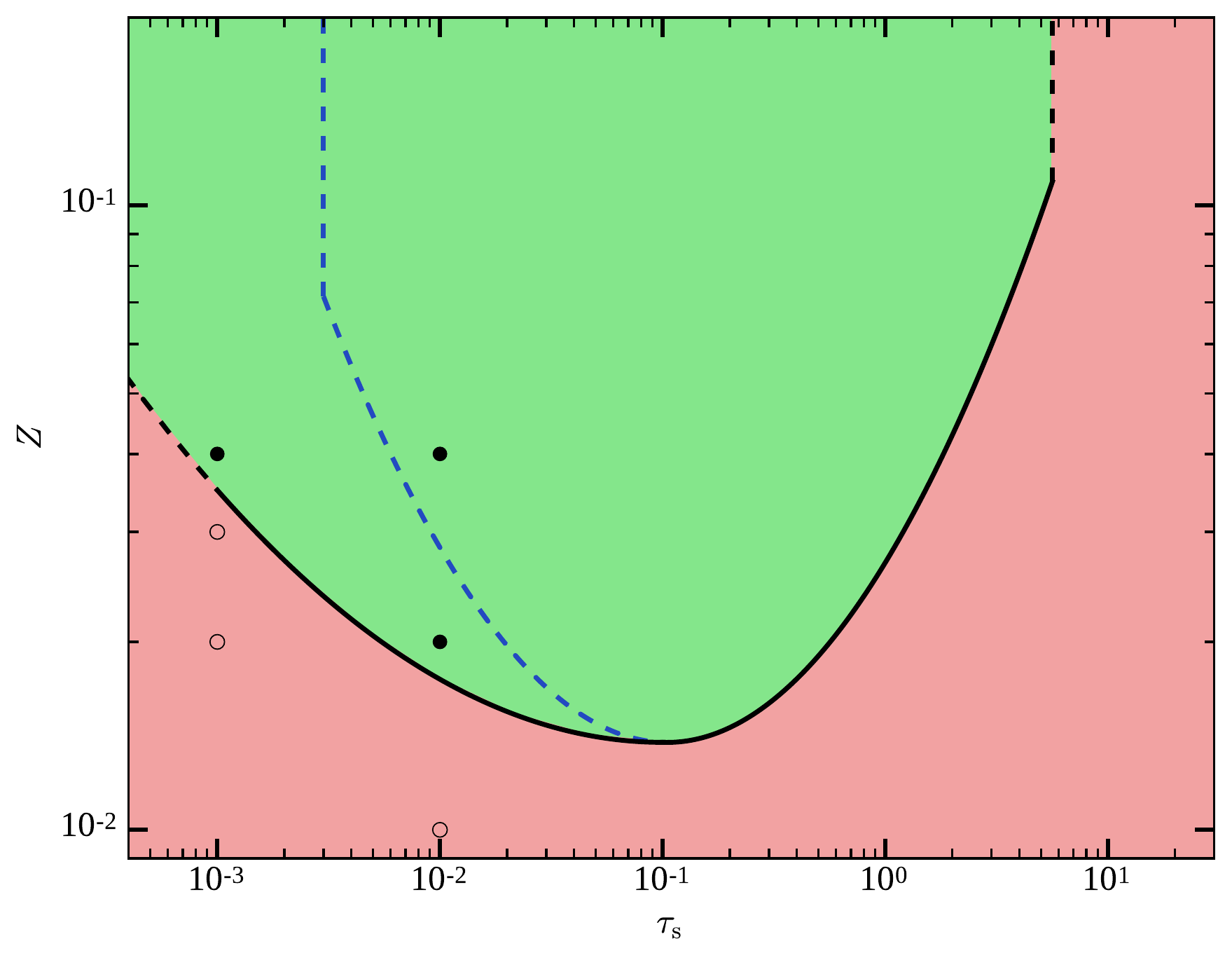}
\caption{Revised critical solid abundance as a function of the dimensionless stopping time $\tau_\mathrm{s}$ from \cite{CJD15}.
The filled and open circles show our models with and without significant radial concentration of solids, respectively.
The solid black line for $\tau_\mathrm{s} \gtrsim 10^{-1}$ and the dashed blue line for $\tau_\mathrm{s} \lesssim 10^{-1}$ are the unmodified critical curve from \cite{CJD15}.
The black solid line for $\tau_\mathrm{s} \lesssim 10^{-1}$ shows the revised part of the critical curve, where the black dashed line is the extrapolation of it.
For solid abundances $Z$ above the black line (green region), spontaneous concentration of solid particles by the streaming instability can occur.
\label{F:Zcrit}}
\end{figure}

Given our findings of a much reduced critical solid abundance $Z_\mathrm{c}$ for which particles of $\tau_\mathrm{s} \ll 1$ can spontaneously concentrate themselves, we hereby revise the $Z_\mathrm{c}$ curve as a function of $\tau_\mathrm{s}$ reported in \cite{CJD15}.
The critical curve for $\tau_\mathrm{s} \gtrsim 0.1$ is the same as in \cite{CJD15} and is given by
\begin{equation}
\log Z_\mathrm{c} = 0.3 (\log\tau_\mathrm{s})^2 + 0.59 \log\tau_\mathrm{s} - 1.57\quad(\tau_\mathrm{s} > 0.1),
\end{equation}
where $\log$ is the logarithm with base 10.
On the other hand, we arbitrarily choose a quadratic function of $\log\tau_\mathrm{s}$ for $\log Z_\mathrm{c}$ such that it passes $Z_\mathrm{c} = 0.035$ at $\tau_\mathrm{s} = 10^{-3}$ and smoothly joins the other curve at $\tau_\mathrm{s} = 0.1$ with a slope of zero.
The resulting function is given by
\begin{equation}
\log Z_\mathrm{c} = 0.10 (\log\tau_\mathrm{s})^2 + 0.20 \log\tau_\mathrm{s} - 1.76\quad(\tau_\mathrm{s} < 0.1).
\end{equation}
The updated critical curve is shown in Fig.~\ref{F:Zcrit}.

Moreover, we note that even though particles of $\tau_\mathrm{s} = 10^{-2}$ can concentrate at a lower solid abundance of $Z = 0.02$, their concentration is one order of magnitude less dense than  particles of $\tau_\mathrm{s} = 10^{-3}$ at $Z = 0.04$.
The maximum local solid density reached by particles of $\tau_\mathrm{s} = 10^{-3}$ is about 300 times the background gas density in the mid-plane $\rho_0$, while that by particles of $\tau_\mathrm{s} = 10^{-2}$ is only about 20$\rho_0$--30$\rho_0$.
By comparing $Z = 0.02$ and $Z = 0.04$ for particles of $\tau_\mathrm{s} = 10^{-2}$ (Sect.~\ref{SS:2dt2}), we note also that with only a difference of twice the solid abundance, the response of the saturated state for particles of $\tau_\mathrm{s} = 10^{-2}$ can be more than one order of magnitude stronger.
At $Z = 0.04$, they reach a peak density of $\sim$100$\rho_0$--300$\rho_0$, on the same order of that by particles of $\tau_\mathrm{s} = 10^{-3}$.
However, due to the formation of relatively more dense filaments of solids at smaller separations for particles of $\tau_\mathrm{s} = 10^{-2}$, the saturated peak density remains comparatively lower than that of particles of $\tau_\mathrm{s} = 10^{-3}$.

This observation leads to an interesting scenario for planetesimal formation.
Solid particles of smaller sizes might be more predisposed towards gravitational collapse inside self-induced dense filaments of solids than those of larger sizes.
To see this, the Roche density $\rho_\mathrm{R}$ is given by
\begin{align}
\frac{\rho_\mathrm{R}}{\rho_0}
&= \frac{9M_\star}{4\pi a^3\rho_0}\\
&= 270\left(\frac{M_\star}{M_\sun}\right)
      \left(\frac{a}{2.5\textrm{ au}}\right)^{-3}
      \left(\frac{\rho_0}{10^{-10}\textrm{ g cm}^{-3}}\right)^{-1},
\end{align}
where $M_\star$ and $a$ are the mass of and the distance to the central star, respectively, and $\rho_0$ is the gas density in the mid-plane.
The normalization is applied at the location of the Asteroid Belt in the MMSN.
We note that the observed gas densities in the terrestrial region of a young protoplanetary disk can be one or two magnitudes higher \citep[we refer to e.g.,][and references therein]{DS14}, and a gas density of $10^{-8}$--$10^{-10}$ g cm$^{-3}$ for $a < 5$ au can also be seen in numerical simulations of star-irradiated accretion disk models at their earliest evolution stages \citep{BJ15,BCP16}.
In any case, the solid concentration achieved in our models with particles of $\tau_\mathrm{s} = 10^{-3}$ is well above the Roche density in the terrestrial region of a wide range of disks when the solid abundance $Z \sim 0.04$, while it may require particularly dense disks or further outward locations for particles of $\tau_\mathrm{s} = 10^{-2}$ to drive planetesimal formation unless $Z \gtrsim 0.04$ also.
Most importantly, we have demonstrated that inside the ice line, strong concentration of mm-sized solid particles, whose growth is limited by bouncing and fragmentation \citep{ZO10,BKE12}, and hence planetesimal formation therein is possible, bridging the problematic gap between dust coagulation and planetesimal formation \citep{DD14}.

Despite that, for a disk with an average solid abundance below the critical one, some mechanisms are still required to enhance the abundance in a local but somewhat larger scale for planetesimals to be able to form via the streaming instability.
Some interesting possibilities include photo-evaporation of the gaseous disk \citep{CG17}, radial pile-up of solids \citep{DAM16,GLM17}, and ice condensation and sublimation \citep{RJ13,IG16,SO17}.

Some discussion on our simplifying assumption of particles of the same size is in order here.
It has been suggested by \cite{BS10a} that the particle-gas dynamics for a population of particles of various sizes is predominantly driven by the largest ones.
They found a critical solid abundance of $Z_c \sim 0.05$ for particles of seven species ranged from $\tau_\mathrm{s} = 10^{-4}$ to $10^{-1}$, and conjectured that it is the total abundance of the particles with $\tau_\mathrm{s} \gtrsim 10^{-2}$ that forms the criterion for triggering strong concentration of particles; that is, $Z(\tau_\mathrm{s} \gtrsim 10^{-2}) \gtrsim 0.02$.
However, if the dust coagulation is limited by the bouncing barrier, the largest particles in the distribution may only have $\tau_\mathrm{s} \sim 10^{-3}$ \citep{ZO10,DD14}.
In this case, the conjecture by \cite{BS10a} breaks down and it is not clear which size range of the particles would dictate the particle-gas dynamics in the system.
On the other hand, when the dust coagulation is limited by the bouncing barrier, the distribution of the particle size tends to concentrate within about one order of magnitude near the barrier \citep{ZO10}.
Therefore, our assumption of particles of the same size might still assimilate this scenario.

Finally, several open questions remain in concentrating small particles by the streaming instability.
We have observed further sedimentation in dense filaments in our models, indicating that the velocity dispersion of the particles decreases in the dense regions.
It remains to be seen if further dust growth could proceed inside such an environment, given the reduced relative velocities, and if the dust growth in turn could drive more spontaneous concentration of solid materials via the streaming instability.
Moreover, the presence of the magnetic fields could render rich structure in the gaseous protoplanetary disks, and the disk mid-plane can be turbulent with various strengths at different locations \citep{TF14}, which significantly affects the ability of solid materials to sediment onto the mid-plane \citep[e.g.,][]{OH11,ZSB15}.
This has nontrivial effects on the dust coagulation and the radial concentration of solids for planetesimal formation to occur \citep{JB14,TB14}.

\section{Summary} \label{S:summary}

In this work, we revisit the condition for the spontaneous concentration of solid particles driven by the streaming instability in the regime of dimensionless stopping time $\tau_\mathrm{s} \ll 1$, as originally investigated by \cite{CJD15}.
Using the local-shearing-box approximation, we simulate a sedimented layer of solid particles of the same size in a laminar gaseous environment, developing streaming turbulence and possibly the ensuing radial concentration of particles.
With the assistance of the numerical algorithm for the stiff mutual drag force by \cite{YJ16}, we are able to conduct the same simulation models of \cite{CJD15} with significantly higher resolutions and longer simulation times.
We focus on two stopping times $\tau_\mathrm{s} = 10^{-2}$ and $10^{-3}$, and systematically vary the solid-to-gas column density ratio $Z$ and the resolution, with two computational domains of $0.2H \times 0.2H$ and $0.4H \times 0.4H$ in the radial-vertical plane, where $H$ is the local scale height of the gas.
We also conduct a few 3D counterparts of the same models and find similar results.

We find that small solid particles can indeed spontaneously concentrate themselves into high density via the streaming instability, given enough time and resolution.
For particles of $\tau_\mathrm{s} = 10^{-2}$ and $10^{-3}$, the critical solid abundance $Z_\mathrm{c}$ above which strong concentration of solids occurs is in the range of $0.01 < Z_\mathrm{c} < 0.02$ and $0.03 < Z_\mathrm{c} < 0.04$, respectively.
The timescales required for this process to reach saturation are approximately 400$P$--1000$P$ and 600$P$--2000$P$, respectively, where $P$ is the local orbital period.
For $\tau_\mathrm{s} = 10^{-2}$ at a solid abundance of $Z = 0.02$, a maximum local solid density of 20$\rho_0$--30$\rho_0$ is observed, where $\rho_0$ is the initial mid-plane density of the gas, while $\tau_\mathrm{s} = 10^{-3}$ at a solid abundance of $Z = 0.04$, results in a density of $\sim$300$\rho_0$.
We also find a super-linear dependence of the solid concentration on the solid abundance.
With these new measurements, we hereby revise the critical solid abundance curve of \cite{CJD15} in the regime of $\tau_\mathrm{s} \ll 1$ in Fig.~\ref{F:Zcrit}.
For a solid density of $\gtrsim$10$^2 \rho_0$, direct gravitational collapse of mm-sized particles can occur and form planetesimals in the terrestrial region of typical protoplanetary disks, as long as the solid abundance reaches the level of a few percent.
This may help alleviate the problematic size gap between the largest particles from dust coagulation and the smallest particles that can spontaneously concentrate themselves via the streaming instability, especially inside the ice line of a protoplanetary disk.

\begin{acknowledgements}

We thank the anonymous referee, Tristan Guillot, and Joanna Dr{\c a}{\.z}kowska for their constructive comments, which helped to significantly improve the manuscript.
All the simulation models presented in this work were performed on resources provided by the Swedish National Infrastructure for Computing (SNIC) at PDC Center for High Performance Computing in KTH Royal Institute of Technology and at Lunarc in Lund University, both of which are located in Sweden.
This research was supported by the European Research Council under ERC Starting Grant agreement 278675-PEBBLE2PLANET.
A.J.\ is grateful for financial support from the Knut and Alice Wallenberg Foundation and from the Swedish Research Council (grant 2010-3710).
\end{acknowledgements}
\bibliographystyle{aa}
\bibliography{30106JN.redline}
\begin{appendix}

\section{Particle resolution} \label{S:parres}

The solid particles in a local patch of the protoplanetary disk are too numerous to be evolved individually in current numerical simulations.
Therefore, the approach of the so-called super-particles or representative particles is often adopted.
Each super-particle models an arbitrarily high number of identical physical particles and assumes the collective dynamical properties of them.
The more super-particles are used, the more the system of physical particles is resolved.

On the other hand, under the particle-mesh construct, the numerical accuracy of treating any interaction between the gas and the particles is limited by the resolution of the fixed grid \citep[e.g.,][]{YJ07}, and hence advantage may not necessarily be gained by a great number of particles within one grid cell.
In the context of the nonlinear evolution of the streaming instability without vertical gravity, it has been shown that the results are insensitive to the number of super-particles used \citep{JY07,BS10b}.
In particular, \cite{BS10b} showed that the resolution of the grid is appreciably more important in obtaining a convergent density distribution of particles, and one particle per cell on average is sufficient in reproducing the distribution.

\begin{figure}
\centering
\includegraphics[width=\columnwidth]{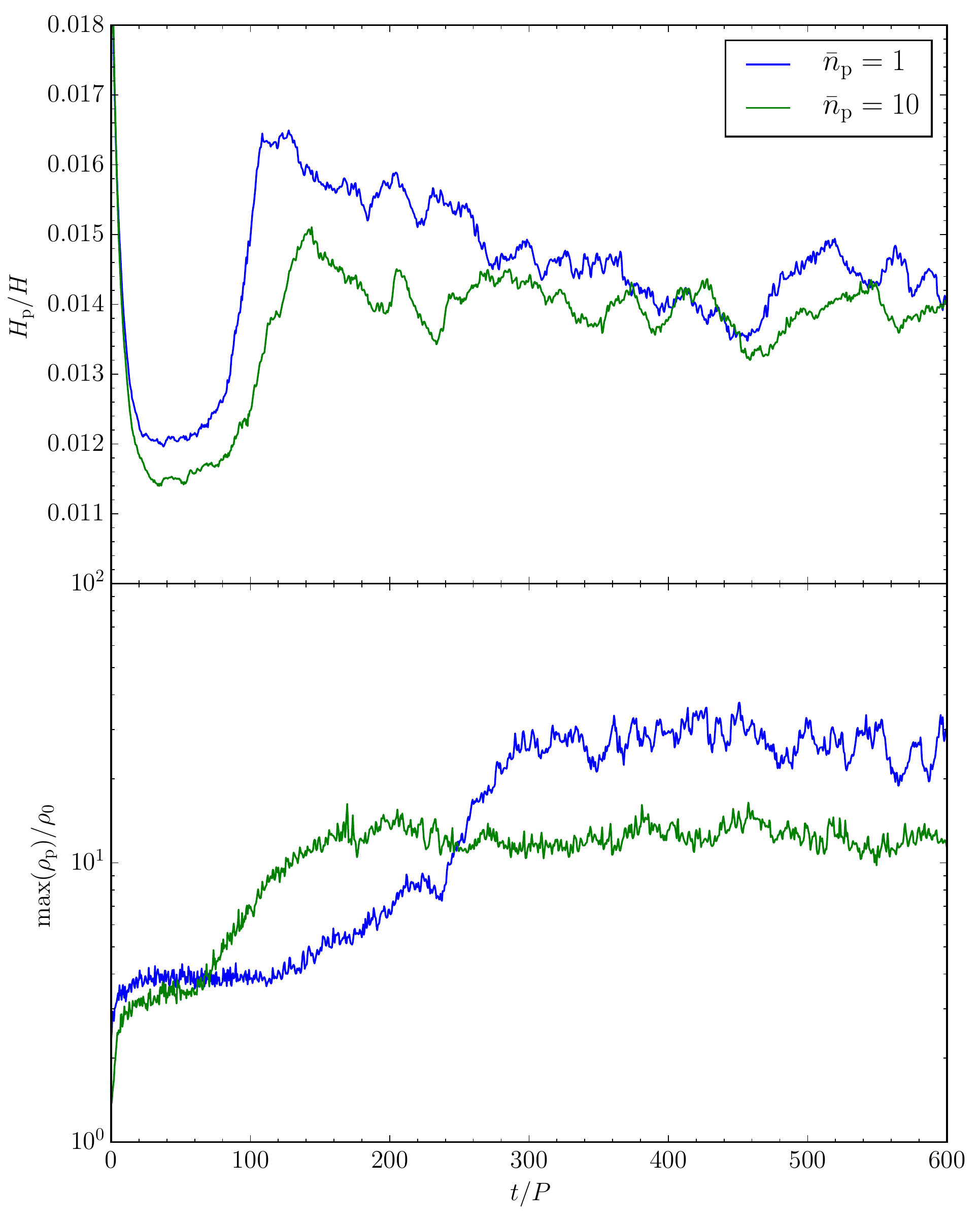}
\caption{Scale height of the particle layer $H_\mathrm{p}$ (top panel) and maximum of the local particle density $\rho_\mathrm{p}$ (bottom panel) as a function of time for two 2D models with different numbers of super-particles.
Both models have particles of dimensionless stopping time $\tau_\mathrm{s} = 10^{-2}$, a solid abundance of $Z = 0.02$, a computational domain of $0.2H \times 0.2H$, and a resolution of 2560$H^{-1}$.
The blue and green lines represent an average number of particles per cell of $\bar{n}_\mathrm{p} = 1$ and 10, respectively.
\label{F:tsnp}}
\end{figure}

To investigate the potential effects of the particle resolution on our results, we select one of our 2D models presented in Sect.~\ref{SS:2dt2} and simulate the same model with ten times more super-particles, that is, with an average number of particles per cell of $\bar{n}_\mathrm{p} = 10$ (see Sect.~\ref{S:method}).
The model has particles of dimensionless stopping time $\tau_\mathrm{s} = 10^{-2}$, a solid abundance of $Z = 0.02$, a computational domain of $0.2H \times 0.2H$, and a resolution of 2560$H^{-1}$.
Fig.~\ref{F:tsnp} compares the scale height of the particle layer and the maximum local particle density as a function of time for this model with $\bar{n}_\mathrm{p} = 1$ and 10 (c.f., Fig.~\ref{F:ts2d}).
The simulation with $\bar{n}_\mathrm{p} = 10$ follows similar evolution as the one with $\bar{n}_\mathrm{p} = 1$: Initial sedimentation, saturation of the streaming turbulence, excitation of the particle layer, and finally radial concentration of the solid particles.
The former reaches the saturated state somewhat earlier than the latter ($t \sim 200P$ versus $t \sim 400P$) and has the same scale height of $(0.0140 \pm 0.003)H$.
On the other hand, the simulation with $\bar{n}_\mathrm{p} = 10$ obtains a saturated maximum local solid density of $(12 \pm 1)\rho_0$, roughly a factor of two smaller than that of the simulation with $\bar{n}_\mathrm{p} = 1$.
The radial and vertical velocity dispersions are also slightly lower at $\Delta v_{\mathrm{p},x} = (0.0027 \pm 0.0001)c_\mathrm{s}$ and $\Delta v_{\mathrm{p},z} = (0.0028 \pm 0.0001)c_\mathrm{s}$, respectively (c.f., Table~\ref{T:sat2d}).

\begin{figure}
\centering
\includegraphics[width=\columnwidth]{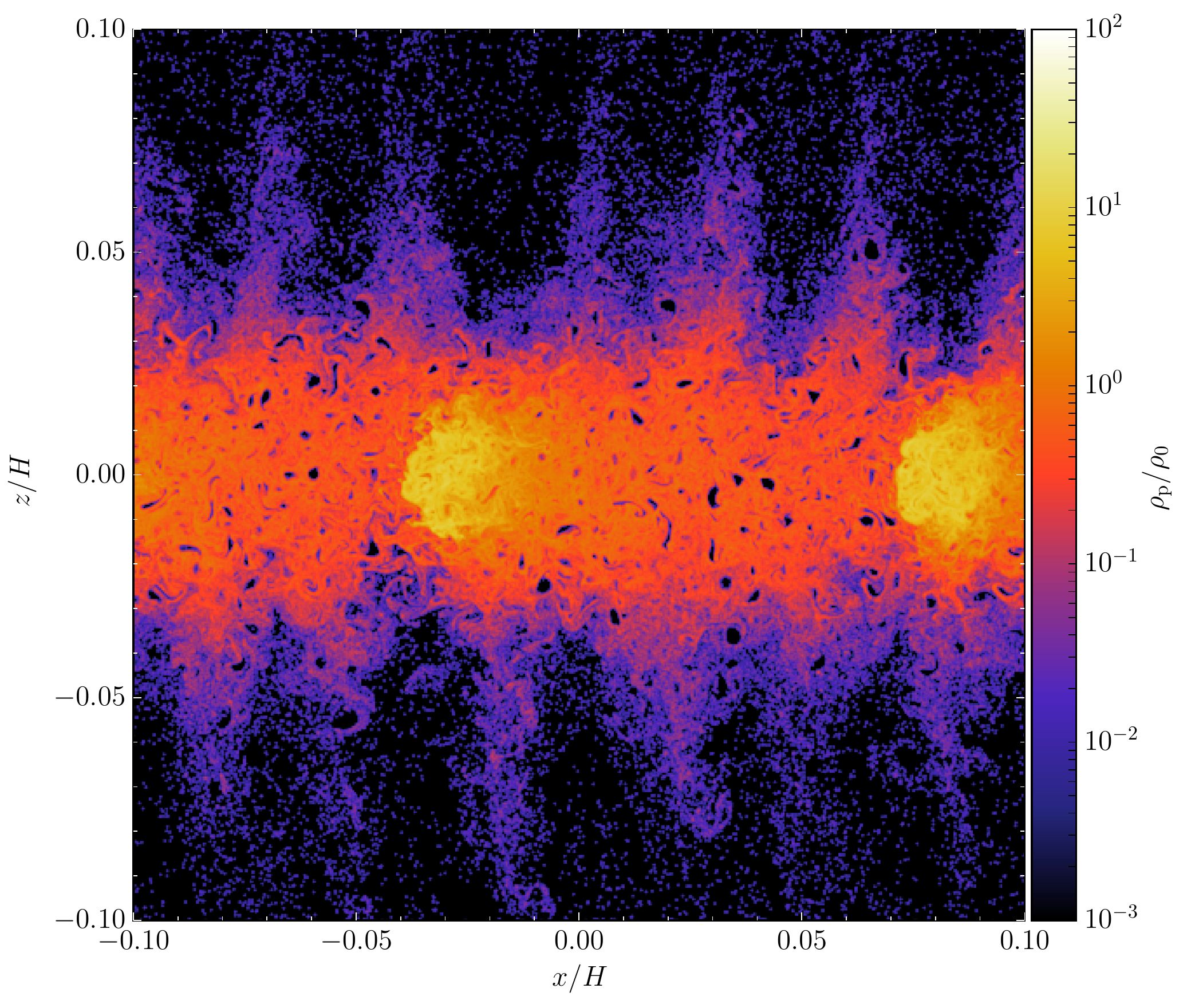}
\caption{Side-view of the particle density at time $t = 300P$ from the 2D model of Fig.~\ref{F:tsnp} with an average number of particles per cell of $\bar{n}_\mathrm{p} = 10$.
\label{F:rhopnp10}}
\end{figure}

The reason for the smaller maximum solid density in the simulation with $\bar{n}_\mathrm{p} = 10$ is that two well-separated axisymmetric filaments of solids are formed in the computational domain, as shown in Fig.~\ref{F:rhopnp10} (c.f., Fig.~\ref{F:evolt2}).
We note that the simulation with $\bar{n}_\mathrm{p} = 1$ also exhibits a transient stage of two coexisting filaments around $t \sim 240P$ before one final dominant filament emerges (Fig.~\ref{F:sigpt2}).
As demonstrated in Figs.~\ref{F:sigp2d}, nonlinear interaction between adjacent filaments may or may not occur in different systems, and the final number of dominant filaments may be uncertain by one or two.
Therefore, the maximum local density of particles can be uncertain within a factor of a few in the saturated state of radial concentration of solid materials.

\pagebreak
\section{Radial pressure gradient in the mid-plane} \label{S:rpg}

As far as the back reaction from the solid particles to the gas via the mutual drag force is concerned, interesting dynamics may occur when the local solid-to-gas density ratio $\epsilon \equiv \rho_\mathrm{p} / \rho_\mathrm{g} \gtrsim 1$.
In this case, while solids tend to radially drift towards the star, they have significant collective momentum to drive the gas to move radially outwards by the conservation of angular momentum.
It has been suggested that a local pressure bump in the gas may form via this mechanism when $\epsilon \sim 1$ and $\tau_\mathrm{s} \sim 1$ \citep{GLM17}.

\begin{figure}[!ht]
\centering
\begin{subfigure}{\columnwidth}
   \centering
   \includegraphics[width=\textwidth]{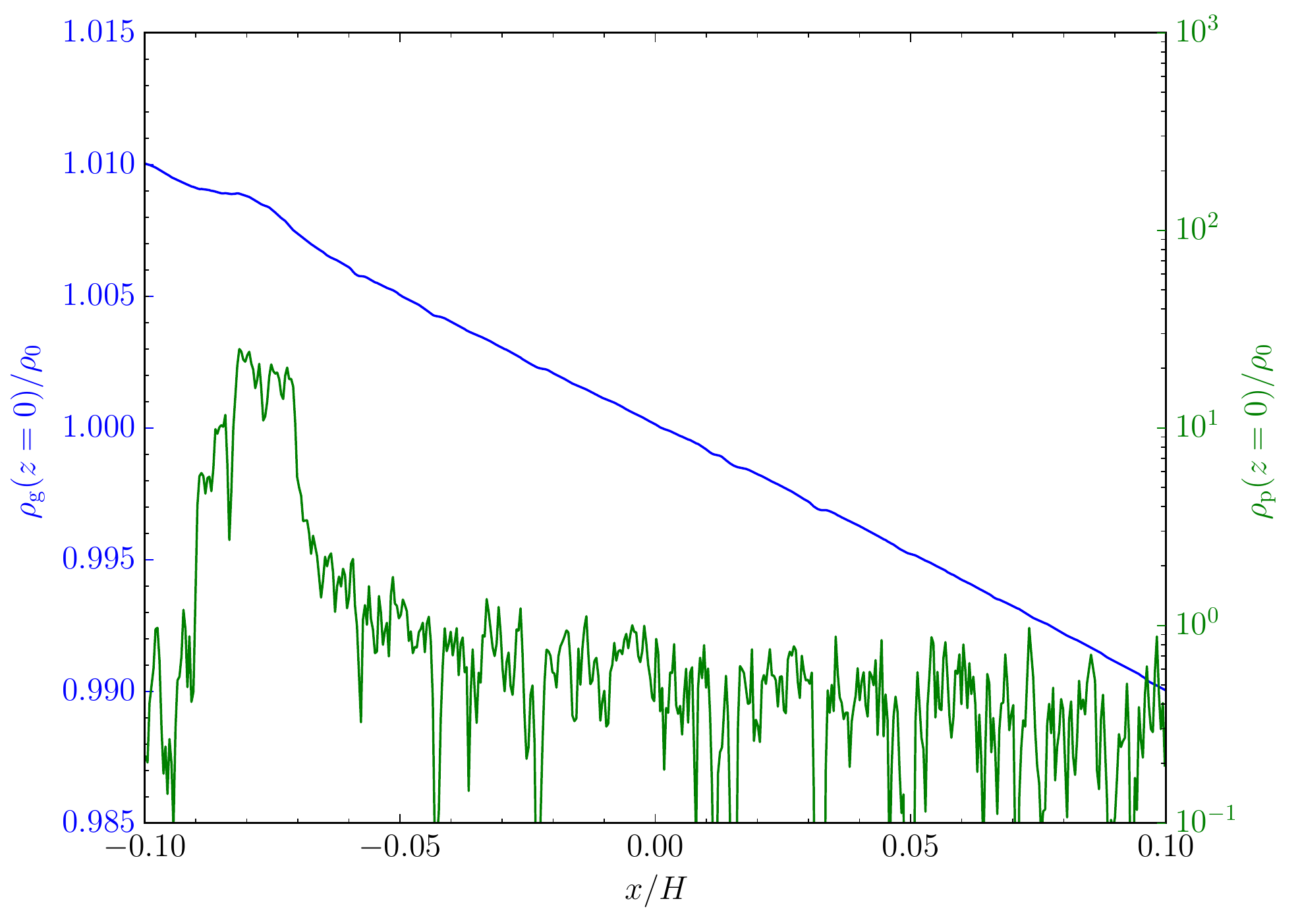}
   \caption{$\tau_\mathrm{s} = 10^{-2}$, $Z = 0.02$, $2560H^{-1}$, $t = 1000P$}
   \label{F:mprho1}
\end{subfigure}
~
\begin{subfigure}{\columnwidth}
   \centering
   \includegraphics[width=\textwidth]{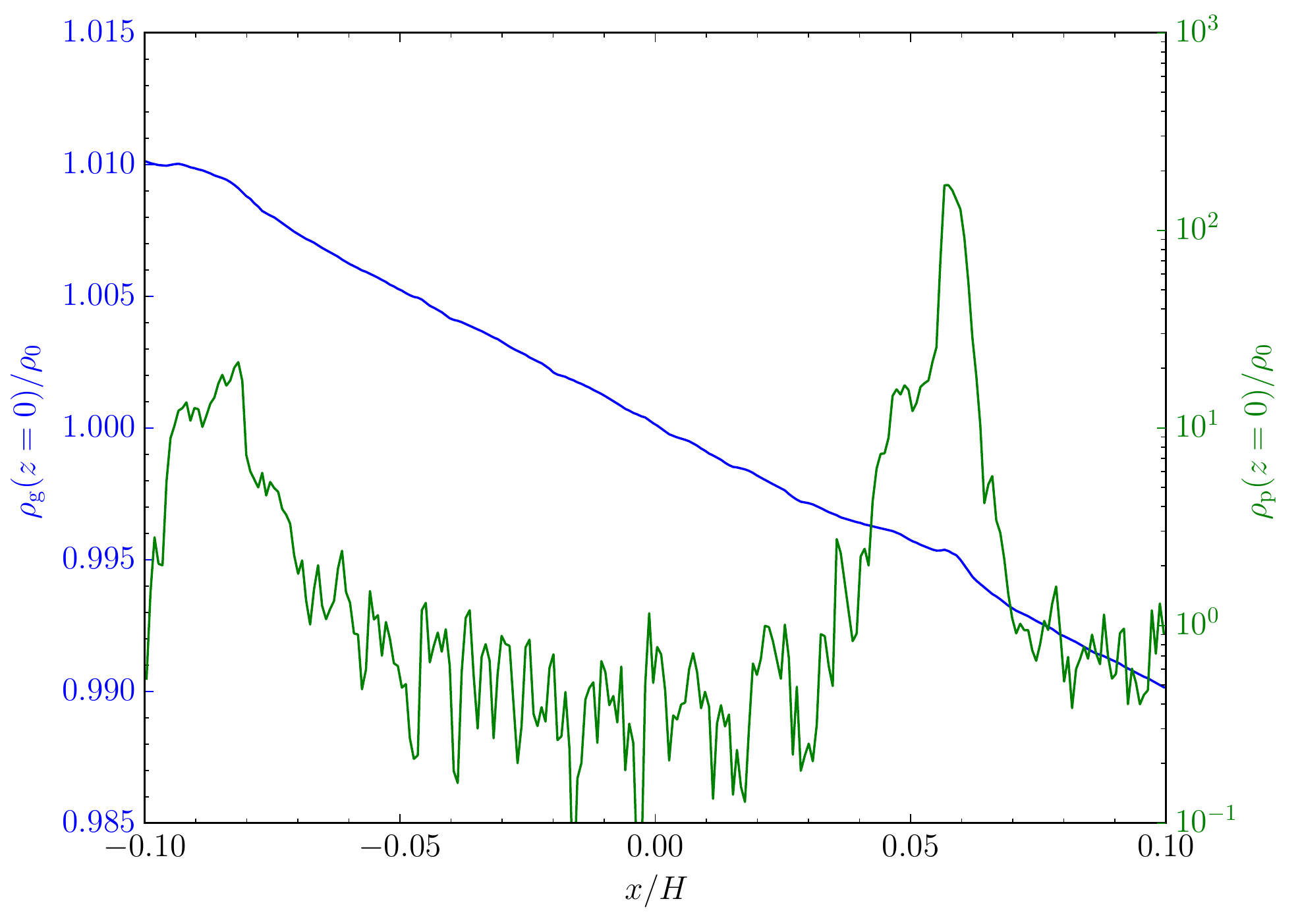}
   \caption{$\tau_\mathrm{s} = 10^{-2}$, $Z = 0.04$, $1280H^{-1}$, $t = 1000P$}
   \label{F:mprho2}
\end{subfigure}
~
\begin{subfigure}{\columnwidth}
   \centering
   \includegraphics[width=\textwidth]{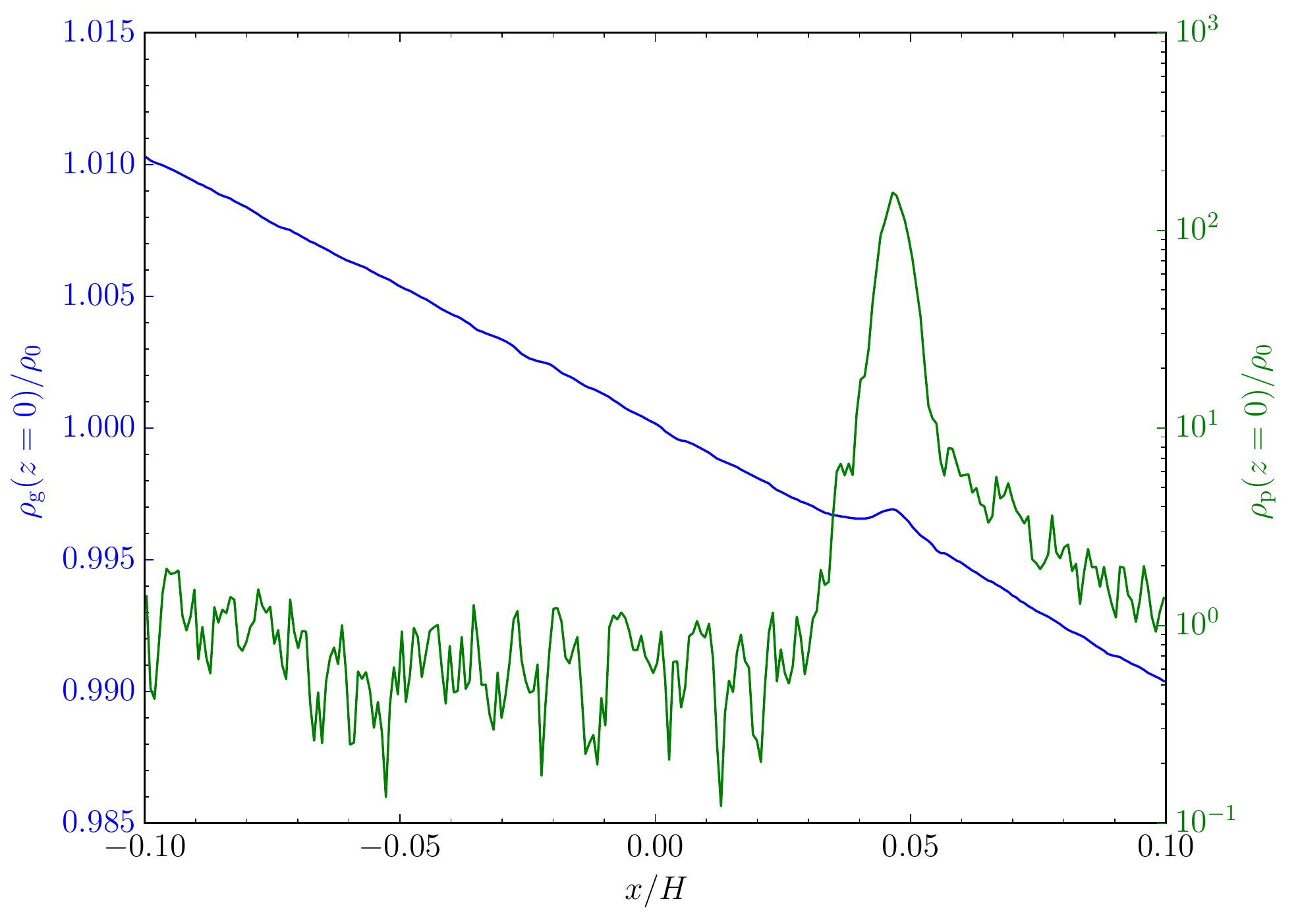}
   \caption{$\tau_\mathrm{s} = 10^{-3}$, $Z = 0.04$, $1280H^{-1}$, $t = 2500P$}
   \label{F:mprho3}
\end{subfigure}
\caption{Final density profiles of gas (blue, left axis) and particles (green, right axis) in the mid-plane for various 2D models with the dimensions of $0.2H \times 0.2H$.
The background density gradient is added to the gas profile for easy recognition of any pressure bump.
\label{F:mprho}}
\end{figure}

In Figs.~\ref{F:mprho}, we plot the final density profiles of the gas and the particles in the mid-plane for several of our 2D models, in which strong local concentration of solids occurs.
The background density gradient $-\left(2\Pi\Omega_\mathrm{K} / c_\mathrm{s}\right)\rho_0 x$ (see Sect.~\ref{S:method}) is added to the gas profile for easy recognition of any pressure bump in our systems.
In all our models, the solid-to-gas density ratio in the mid-plane reaches $\epsilon \sim 1$ simply due to the process of sedimentation.
For particles of $\tau_\mathrm{s} = 10^{-2}$, we find that the perturbation in the gas density is in general small with respect to the background density gradient unless $\epsilon \gtrsim 10$.
Even for $\epsilon \sim 10$--100, a pressure bump barely forms near a local density peak of solids (Figs.~\ref{F:mprho1} and~\ref{F:mprho2}).
On the other hand, it appears that a local pressure bump that coincides with the local density peak of solids does readily occur for particles of $\tau_\mathrm{s} = 10^{-3}$ when $\epsilon \gg 1$ (Fig.~\ref{F:mprho3}).
Therefore, it seems that in our systems, $\tau_s \ll 1$ and $\epsilon \gg 1$ is required to generate dynamically significant perturbation in the gas via the back reaction of the drag force.

\section{Non-axisymmetric particle-gas dynamics in three dimensions} \label{S:nonaxi}

The 3D models presented in Sect.~\ref{S:3d} exhibit an interesting phenomenon that does not occur in the 2D models described in Sect.~\ref{S:2d}.
The dense filaments of solid materials appear to migrate radially outwards, as shown in Fig.~\ref{F:sigp3d}.
In this appendix, we discuss the particle-gas dynamics associated with this phenomenon in more detail.

\begin{figure}
\centering
\includegraphics[width=\columnwidth]{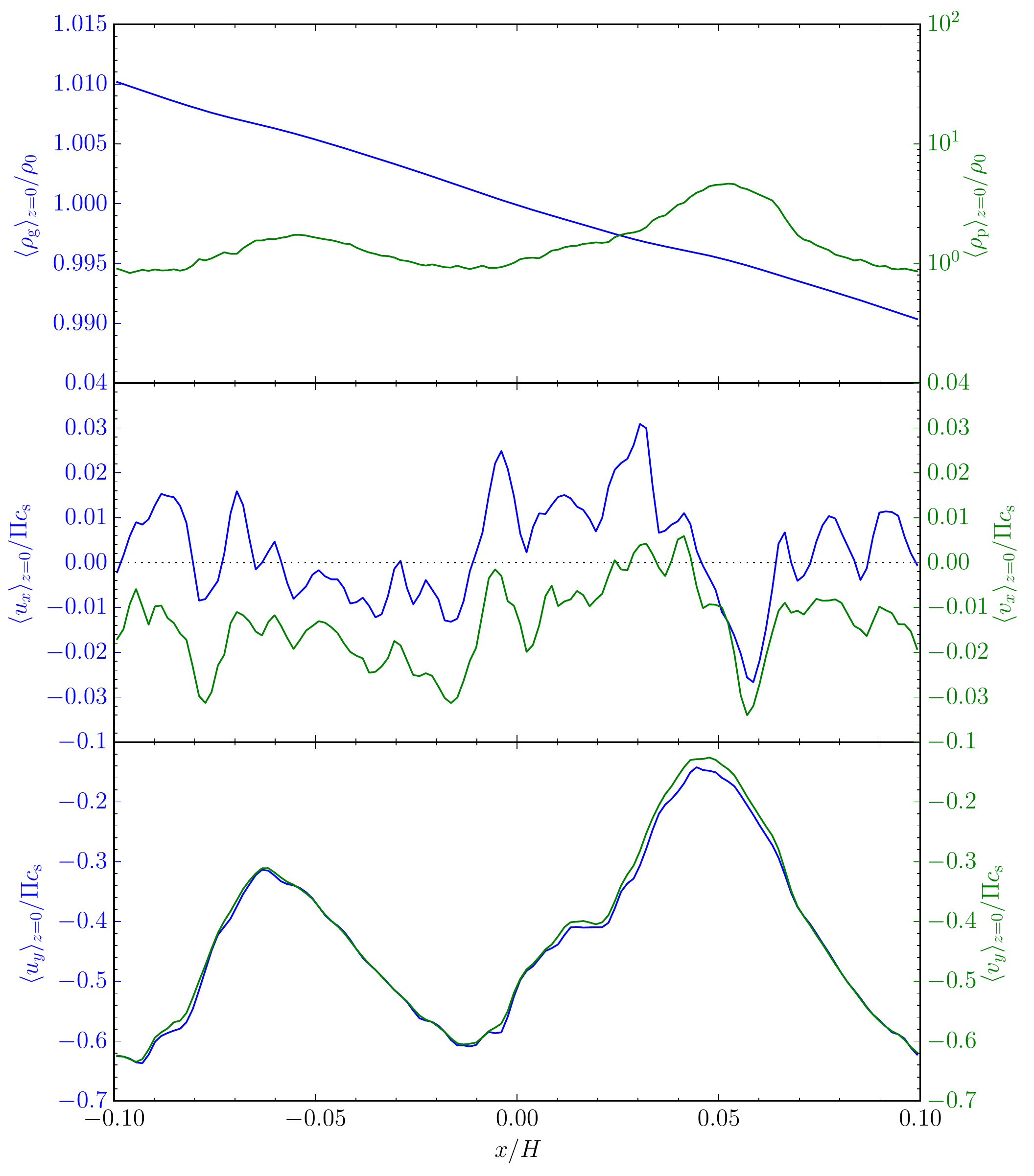}
\caption{Azimuth-averaged radial profiles of the particle/gas properties in the mid-plane at $t = 600P$ for the 3D model with particles of $\tau_\mathrm{s} = 10^{-2}$ and a resolution of 640$H^{-1}$.
The top, middle, and bottom panels show densities, radial velocities, and azimuthal velocities with respect to the background Keplerian shear, respectively.
The blue (left axis) and the green (right axis) lines represent the gas and the solid particles, respectively.
The background density gradient is added to the gas profile for easy recognition of any pressure bump (see Appendix~\ref{S:rpg}).
The gas and the particle densities are normalized by the initial gas density in the mid-plane $\rho_0$, while the velocities are weighted by the densities and normalized by the velocity reduction of the gas $\Delta v = \Pi c_\mathrm{s}$ due to external radial pressure gradient (Sect.~\ref{S:method}).
We note that a zero velocity is equivalent to moving at the Keplerian velocity, irrespective of the location.
\label{F:mprprof}}
\end{figure}

Fig.~\ref{F:mprprof} shows the azimuth-averaged radial profiles of the particle/gas properties in the mid-plane at $t = 600P$ for the 3D model with particles of dimensionless stopping time $\tau_\mathrm{s} = 10^{-2}$ and a resolution of 640$H^{-1}$ (c.f., Figs.~\ref{F:evol3dt2} and~\ref{F:sigp3d}).
The dense filament of solids at this moment is located at $x \sim +0.05H$.
Due to the tight coupling between the gas and the particles, the relative velocity between them are small with respect to the speed of sound $c_\mathrm{s}$.
The azimuthal velocities of the gas and the particles near the filament remain sub-Keplerian.
More importantly, the radial velocities of the particles in the filament are predominantly inward, on the order of $-0.02 \Pi c_\mathrm{s}$.
For comparison, the outward radial velocity of the filament shown in the left panel of Fig.~\ref{F:sigp3d} is about $+4\times10^{-4} \Pi c_\mathrm{s}$.
On the other hand, the gas density profile near the filament indicates a weak density bump.
However, the magnitude of the radial pressure gradient this bump generates is appreciably less than the external one.
In other words, this density bump is not a dynamically significant one.
Therefore, we conclude that the apparent outward radial migration of the dense filament of solids is a pattern movement, instead of a material one.

\begin{figure}
\centering
\begin{subfigure}{\columnwidth}
   \centering
   \includegraphics[width=\textwidth]{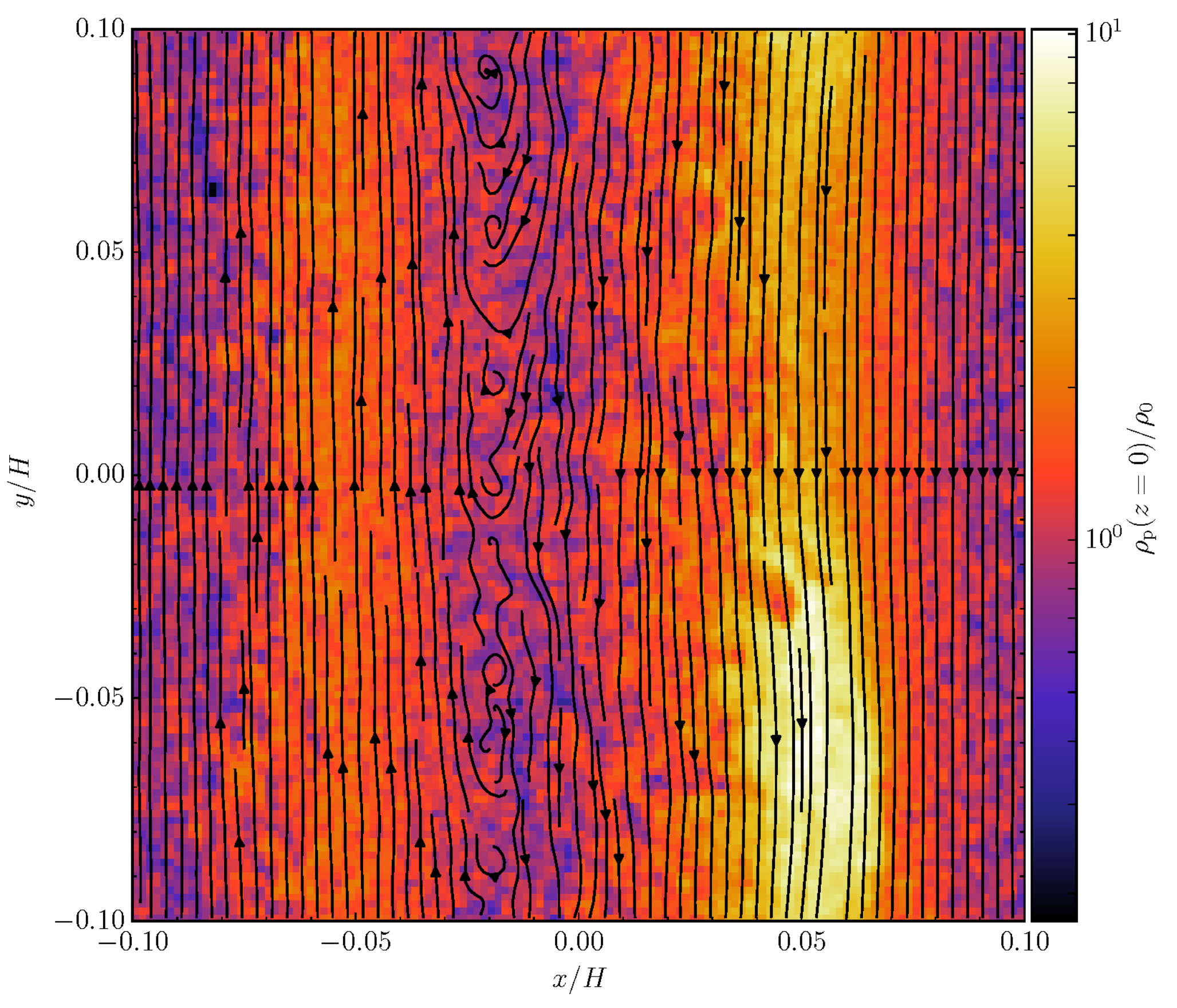}
   \caption{particles kinematics}
\end{subfigure}
~
\begin{subfigure}{\columnwidth}
   \centering
   \includegraphics[width=\textwidth]{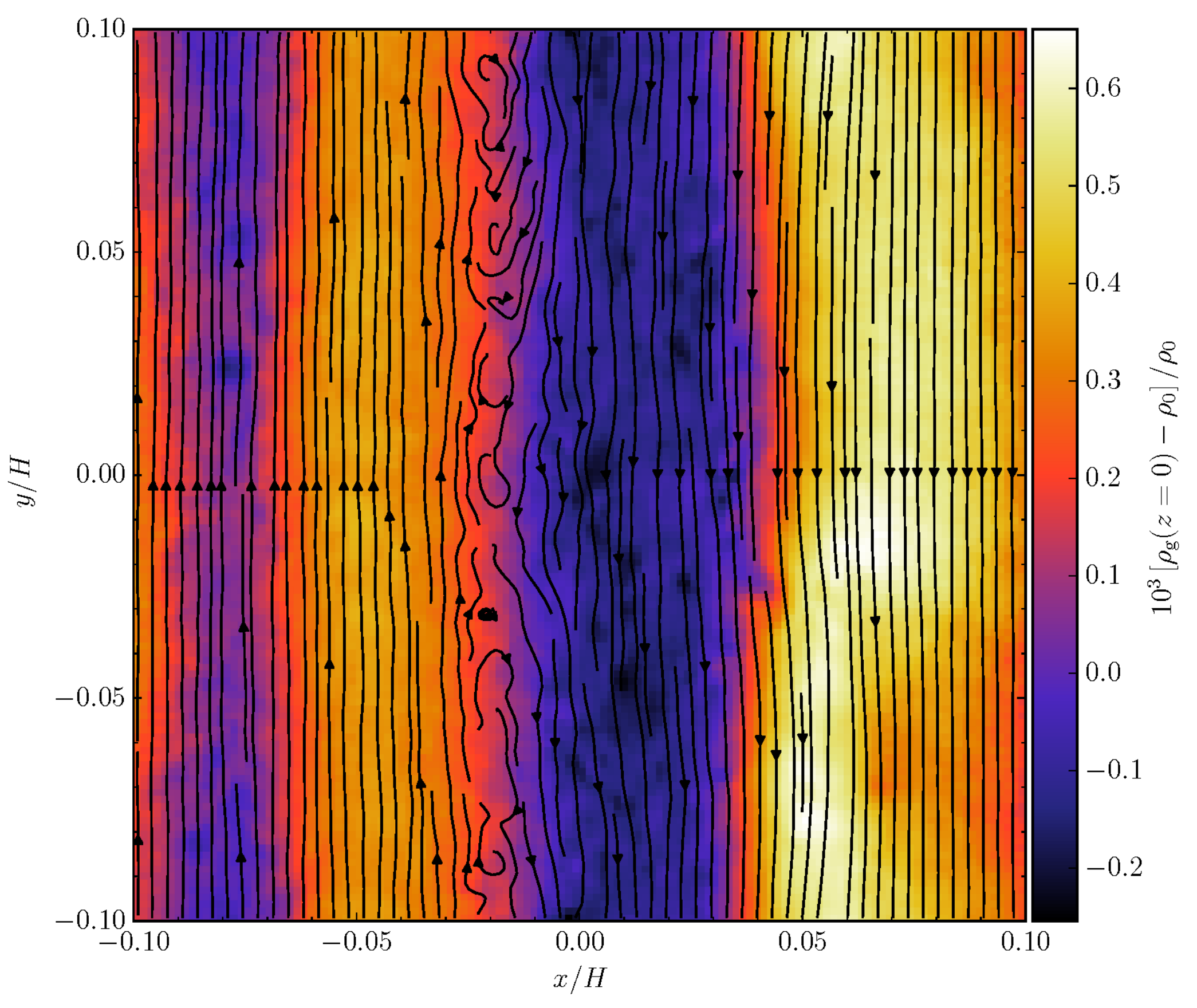}
   \caption{gas kinematics}
\end{subfigure}
\caption{Streamlines in the mid-plane at $t = 600P$ for the 3D model with particles of $\tau_\mathrm{s} = 10^{-2}$ and a resolution of 640$H^{-1}$ for (a)~the particles and (b)~the gas.
The background shows the densities of the particles and the gas, respectively.\label{F:mpkin}}
\end{figure}

The outward pattern movement of the filament of solids seems to result from the large-scale non-axisymmetric structure of the particle-gas system, which is absent in our 2D models.
As shown in Fig.~\ref{F:evol3dt2}, the solid particles concentrate not only radially, but also azimuthally.
Fig.~\ref{F:mpkin} shows a slice through the mid-plane at the same time from the same model as in Fig.~\ref{F:mprprof}.
Also shown are the streamlines of the gas and the particles.
We note that the streamlines are slightly bent radially outwards near the concentration of the particles as well as the gas.
This indicates that the particles azimuthally above the concentration tend to move outwards while those below tend to move inwards.
The less dense region at the opposite azimuthal phase display an opposite trend, that is, the streamlines are slightly bent radially inwards.
We also note that the radially depleted region between the filaments of solids signals small-scale vortex motions.
The mechanism that leads to this type of non-axisymmetric structure and movement remains unclear and requires future dedicated investigation.

Similar pattern of the particle-gas dynamics described above is also manifested in the 3D models with particles of $\tau_\mathrm{s} = 10^{-3}$.
However, as shown in Fig.~\ref{F:evol3dt3} and the right panel of Fig.~\ref{F:sigp3d}, once the apparent radially outward movement of the two weak filaments joins the third stronger one, an axisymmetric dense filament of solids is produced.
The filament appears to be almost stationary afterwards, which is consistent with the particle-gas drag equilibrium solution of \cite{NSH86}.
This further supports the idea that the apparent radially outward movement of filaments is exclusively a non-axisymmetric phenomenon.

\end{appendix}
\clearpage
\end{CJK}
\end{document}